\documentclass[aps,pra,preprint,amsmath,amssymb,superscriptaddress,longbibliography]{revtex4-2}%投稿预印版（投稿时的统一排版）
\usepackage{graphicx}% 插入图片
\usepackage{url}
\usepackage{booktabs}
\usepackage{multirow}
\usepackage{makecell}
\usepackage{textcomp}
\usepackage{phonetic}
\usepackage{extarrows}
\usepackage{color}
\usepackage{float}% 引入 float 宏包
\usepackage{hyperref} % 如果您的文档类自带了hyperref，这行可以删掉；如果报错说没有hyperref，则保留这行但不要加方括号选项

\hypersetup{
	colorlinks=true,      % 激活彩色链接
	linkcolor=blue,       % 内部链接（如引用编号）为蓝色
	citecolor=blue,       % 文献引用为蓝色
	urlcolor=blue,        % 网址链接为蓝色
	pdfborder={0 0 0}     % 去掉链接周围的方框
}
\usepackage{orcidlink} 
\usepackage{epstopdf}
\usepackage[export]{adjustbox}
\begin{document}
	\title{Resonant Multiphoton Conversion Mediated by a Shared Vibration in Cavity QED}% working title
	
		\author{Yu-Mu Liu}
	\email{Corresponding author: liuym@ahstu.edu.cn}
	\affiliation{School of Electrical and Electronic Engineering, Anhui Science and Technology University, Bengbu, Anhui 233000, China}
	
	\author{Qiu-Yun Yang}
	\affiliation{School of Electrical and Electronic Engineering, Anhui Science and Technology University, Bengbu, Anhui 233000, China}
	
	\author{Na Wang}
	\affiliation{School of Electrical and Electronic Engineering, Anhui Science and Technology University, Bengbu, Anhui 233000, China}
	
	\author{Ming-Lei Guo}
	\email{Corresponding author: Guomlnj@163.com}
	\affiliation{School of Electrical and Electronic Engineering, Anhui Science and Technology University, Bengbu, Anhui 233000, China}
	
	\author{Hong-Fu Wang}
	\email{Corresponding author: hfwang@ybu.edu.cn}
	\affiliation{Department of Physics, College of Science, Yanbian University, Yanji, Jilin 133002, China}
	\affiliation{Institute of Quantum Science and Technology, Yanbian University, Yanji, Jilin 133002, China}
	\affiliation{Institute for Interdisciplinary Quantum Information Technology, Jilin Engineering Normal University, Changchun, Jilin 130052, China}
	\affiliation{Jilin Engineering Laboratory for Quantum Information Technology, Changchun, Jilin 130052, China}
	
	\begin{abstract}
		
	We show that a vibration shared by two emitter subensembles can jointly establish multiphoton resonance and constructive interference among the virtual exchange pathways responsible for coherent conversion. For opposite displacements with an integer magnitude ratio $r\geq2$, the displacements induced by $r$ excitations in the less displaced subensemble and one in the other cancel exactly, a condition we call displacement neutrality. When the emitter and cavity frequencies match, the resulting composite is resonant with $r+1$ cavity photons, while the vibration-mediated cross interaction makes the conversion pathways add constructively, establishing a direct relation between the displacement ratio and the conversion order. In contrast, independent vibrations can be tuned to the same resonance, but their conversion pathways cancel at leading order, demonstrating that energy resonance alone does not guarantee coherent conversion. For $r=2$, this conversion suppresses three-photon coincidences while leaving the two-photon statistics nearly Poissonian. In the weak-driving limit, interference can also screen the two-photon correlations at all delays from the relative direction of the displacements, whereas the three-photon statistics remain sensitive to it. Because both resonant states leave the vibration undisplaced, their coherence is protected from direct dephasing by the shared vibration, leaving emitter dephasing as the tighter constraint. These results identify relative vibrational displacements as a means of controlling coherent multiphoton conversion and its noise response.
		
	\end{abstract}

	%\keywords{quantum correlations, cavity electromagnonic system, directional quantum squeezing}
	\maketitle
	\section{Introduction}\label{sec.1}
	Optical nonlinearities operating at the single-photon level are central to quantum optics and photonic quantum technologies~\cite{2009NatPhoton687,2008Nature1023,2014NatPhoton685,2013RMP299,2015RMP1379}. Conventional photon blockade arises from an anharmonic excitation ladder that detunes the absorption of a second photon after the first excitation, yielding antibunched light with $g^{(2)}(0)<1$ under coherent driving~\cite{1997PRL1467,2005Nature87}. Such photon-number selectivity can extend to higher excitation manifolds~\cite{2013PRA023809,2008PRL203602}, whose statistics are revealed by higher-order correlations such as $g^{(3)}(0)$~\cite{2014PRA023846,2017PRL133604}. In two-photon blockade, two-photon bunching can coexist with suppressed three-photon events~\cite{2017PRL133604,2019PRA053857}, and three-photon antibunching can also occur with nearly Poissonian second-order statistics, $g^{(2)}(0)\simeq1$~\cite{2017PRA063842}. These distinct correlation signatures show that photon statistics depend on the energies of successive excitation manifolds and on the coherent pathways connecting them~\cite{2020LPR1900279}.
	
	One prominent route to conventional photon blockade is the Jaynes--Cummings anharmonicity of a two-level emitter coupled to a cavity mode. Near resonance, the coupled system forms polariton doublets in successive excitation manifolds, with splittings that grow as the square root of the excitation number~\cite{2005Nature87,2007Nature896,2008NatPhys382,2008Nature315}. This mechanism has been demonstrated with atoms~\cite{2005Nature87}, quantum dots~\cite{2008NatPhys859,2015PRL233601}, and superconducting circuits~\cite{2011PRL243601,2011PRL053602}. In each of these platforms, resolving the anharmonic shifts requires a coupling that is sufficiently strong relative to the cavity and emitter linewidths. For an ensemble of emitters coupled to the cavity through single-photon exchange, the collective light--matter coupling is enhanced, while the low-excitation response approaches that of a harmonic mode as the emitter number increases~\cite{2026PRXQ020349,2022PRR033083}. In resonant multiemitter systems, increasing the number of emitters weakens photon blockade, whereas detuning can reverse this trend~\cite{2019PRL243602}. In the dispersive Tavis--Cummings regime, additional emitter driving has been proposed to recover conventional photon blockade~\cite{2023OE24939}. In this anharmonicity-based route, photon-number selectivity therefore depends on resolving the nonlinear energy shifts between successive excitation manifolds.
	
	A second route to photon-number selectivity relies on quantum interference between excitation pathways. In unconventional photon blockade, coupled modes with a weak nonlinearity reach the two-photon state along two pathways that interfere destructively, strongly suppressing the two-photon amplitude even when the nonlinear shifts are much smaller than the linewidths~\cite{2010PRL183601,2011PRA021802,2017PRA053810}. This interference has been observed with a quantum dot coupled to two orthogonally polarized micropillar modes~\cite{2018PRL043601} and in the microwave domain~\cite{2018PRL043602}. Anharmonicity- and interference-based blockade can occur within the same system~\cite{2020LPR1900279,2023SCPMA66240313,2026SCPMA69240313}: in a cavity containing two inhomogeneous emitters, subradiant dressed states give rise to interference-based blockade alongside polaritonic blockade~\cite{2017PRA011801}. Interference-based selectivity is therefore governed by the relative amplitudes and phases of the excitation pathways, which depend on the couplings and complex energies of their intermediate states.
	
	Vibrational degrees of freedom act on both the energies and the pathways of multiphoton excitation. An excitation that displaces a vibrational coordinate acquires a polaron shift, and the nonadditive displacement energy of several excitations produces an effective photon nonlinearity, as in optomechanical photon blockade~\cite{2011PRL063601,2011PRL063602,2014RMP1391}; the displacement also renormalizes the emitter--cavity exchange through the vibrational overlap~\cite{2002PRB235311,2016JPCM103002}. For an ensemble in which each emitter has its own vibration, sufficiently strong collective cavity coupling suppresses the vibrational reorganization associated with the bright collective excitation~\cite{2016PRL238301,2018ACSPhoton249}. A vibrational mode shared by multiple emitters mediates interactions between them~\cite{2010NatPhys602,2013PRL156402}, as collective phonons do in trapped-ion crystals~\cite{2004PRL207901}, and the coupling to a common mechanical coordinate can change sign with the emitter position through the local strain field~\cite{2014NatNano106}. For two emitter subensembles coupled to the same vibration, the relative sign of their couplings, which we call the configuration, is therefore physical and determines the sign of the vibration-mediated cross interaction; for independent vibrations, it can be removed by redefining the coordinates. Vibrational coupling also causes dephasing and phonon scattering~\cite{2002PRB235311,2002PRB195313,2016JPCM103002,2017NatPhoton521}, while correlations in a shared phonon bath can reduce dephasing at particular emitter separations~\cite{2009PRL146404,2025PRB045303}. These features raise two questions: whether configuration dependence can be screened in the two-photon response while persisting in higher-order correlations, and, more importantly, whether the shared displacement structure can jointly determine the resonance, the coherent conversion, and the vibrational-noise response of a selected multiphoton process.
	
	Here, we show that the configuration dependence can be screened in the two-photon response, and that the shared displacement structure jointly determines the resonance, the coherent conversion, and the vibrational-noise response of a three-photon process. When interference keeps the state with one excitation in each subensemble empty, the leading-order two-photon correlations of opposite- and same-side displacements coincide at all delays, whereas their three-photon statistics differ, and two-photon blockade occurs only for opposite displacements over a finite range of couplings. More importantly, when the exchange couplings are weak compared with the displacement energies, opposite displacements with a magnitude ratio of two leave the emitter state with two excitations in the less displaced subensemble and one in the other undisplaced, so that this composite state is degenerate with three cavity photons when the emitters are resonant with the cavity. The vibration-mediated cross interaction both cancels the displacement energy of the composite and makes the three exchange pathways of the conversion add constructively, yielding a finite coupling between the three-photon state and the composite. This conversion suppresses $g^{(3)}(0)$ while $g^{(2)}(0)\simeq1$. Same-side displacements leave the composite off resonance at the same detunings; for independent vibrations, detuning restores the resonance, but the conversion pathways cancel at leading order. For opposite displacements with an integer magnitude ratio $r\geq2$, displacement neutrality selects the coherent conversion of $r+1$ photons, whose leading-order strength we obtain in closed form. The displacement neutrality that creates the resonance also protects it: because both resonant states leave the vibration undisplaced, thermal noise of the shared vibration does not dephase them directly and acts only through virtual admixtures of displaced states, whereas emitter dephasing sets the tighter constraint on the suppression of $g^{(3)}(0)$.
	
	The paper is organized as follows. Section~\ref{sec:model} introduces the model and the vibration-mediated interactions. Section~\ref{sec:correlations} analyzes the configuration dependence of successive excitation manifolds, Sec.~\ref{sec:threephoton} the displacement-compensated three-photon resonance, and Sec.~\ref{sec:dephasing} its dephasing. Section~\ref{sec.5} concludes.
	
	\section{Model and vibration-mediated interactions}
	\label{sec:model}
	\begin{figure}%pdflatex --shell-escape -interaction=nonstopmode %.tex
		\centering
		\includegraphics[width=0.75\linewidth]{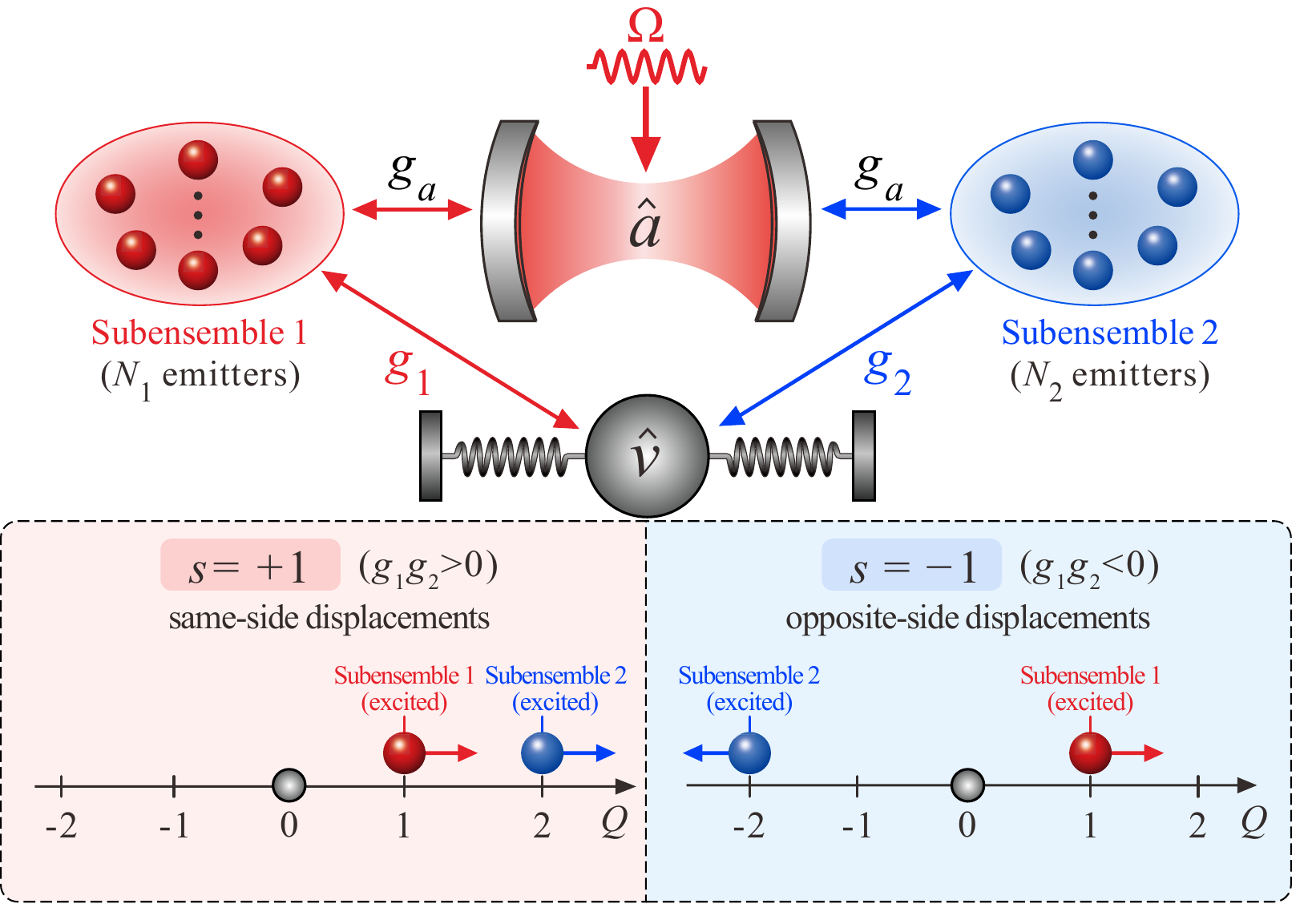}
		\hspace{0in}%
		\caption{Schematic of the system. A single-mode cavity $\hat a$, coherently driven with amplitude $\Omega$, couples with strength $g_a$ to each emitter of two near-degenerate subensembles of $N_1$ and $N_2$ two-level emitters. Both subensembles couple to a common vibrational mode $\hat v$ through the signed couplings $g_1$ and $g_2$, so that every emitter excitation displaces the vibration. Lower panels: displacement $Q$ produced by a single excitation of subensemble 1 or 2, in units of that of subensemble 1, shown for $|g_2|=2g_1$, the smallest of the integer ratios considered in Sec.~\ref{sec:threephoton}. For $s={\rm sgn}(g_1g_2)=+1$, the two subensembles displace the vibration to the same side; for $s=-1$, to opposite sides. The configuration $s$ sets the sign of the vibration-mediated cross interaction $U_{12}=-2g_1g_2/\omega_v$, with $\omega_v$ the vibrational frequency.}
		\label{fig1}
	\end{figure}
	We consider a coherently driven single-mode cavity coupled
	to two near-degenerate subensembles of two-level emitters,
	as illustrated in Fig.~\ref{fig1}. Each subensemble
	($\mu=1,2$) contains $N_\mu$ emitters coupled to the cavity
	mode $\hat a$ with strength $g_a$. Both subensembles also
	interact with a common high-frequency vibrational mode
	$\hat v$, the shared vibration, through excitation-dependent couplings $g_\mu$.
	These signed couplings describe the conditional displacements
	of the shared vibration and distinguish two configurations,
	$s=\mathrm{sgn}(g_1g_2)=\pm1$, corresponding to same-side
	and opposite-side displacements, respectively. 
	
	Within the electric-dipole approximation and the
	rotating-wave approximation for the cavity--emitter
	interaction~\cite{1968PR379}, the Hamiltonian in the laboratory frame
	reads ($\hbar=1$)
	\begin{align}
		\hat H_{\rm lab}(t)={}&
		\omega_a\hat a^\dagger\hat a
		+\omega_v\hat v^\dagger\hat v
		+\sum_{\mu=1}^{2}\sum_{j=1}^{N_\mu}
		\left[
		\omega_\mu
		+g_\mu(\hat v^\dagger+\hat v)
		\right]
		\hat\sigma^+_{\mu j}\hat\sigma^-_{\mu j}
		\nonumber\\
		&+g_a\sum_{\mu=1}^{2}\sum_{j=1}^{N_\mu}
		\left(
		\hat a^\dagger\hat\sigma^-_{\mu j}
		+\hat\sigma^+_{\mu j}\hat a
		\right)	+\Omega\left(
		\hat a^\dagger e^{-i\omega_L t}
		+\hat a e^{i\omega_L t}
		\right),
		\label{eq:Hlab}
	\end{align}
	where $\omega_a$, $\omega_v$, and $\omega_\mu$
	denote the cavity, vibrational, and emitter transition
	frequencies, respectively.
	The operators $\hat\sigma^+_{\mu j}$ and
	$\hat\sigma^-_{\mu j}$ raise and lower the $j$th emitter in subensemble $\mu$, respectively. The cavity is coherently driven with amplitude $\Omega$ at frequency $\omega_L$. The signed coupling $g_\mu$ is the shift of the transition frequency of subensemble $\mu$ per zero-point displacement of the shared coordinate~\cite{2002PRB235311} and determines the direction and magnitude of the excitation-induced equilibrium displacement; $s=-1$ thus describes two subensembles whose transition frequencies shift in opposite directions when the coordinate moves. For emitters coupled to a mechanical mode through strain~\cite{2014NatNano106,2013PRL156402}, the local strain sets both the sign and the magnitude of each coupling; subensembles located where the strain of the mode has opposite signs, for example on opposite sides of the neutral axis of a flexural mode, realize $s=-1$, and the strain magnitudes at their positions set $|g_2|/g_1$.
	Reversing the orientation of this coordinate changes
	$g_1\to-g_1$ and $g_2\to-g_2$ simultaneously
	and leaves $s=\mathrm{sgn}(g_1g_2)$ unchanged.
	Since $s$ also fixes the sign of the vibration-mediated cross
	interaction $U_{12}=-2g_1g_2/\omega_v$ [Eq.~(\ref{eq:HP})],
	it distinguishes physically different relative-displacement
	configurations.
	In contrast, for two independent vibrational modes,
	the sign of each coupling can be changed separately by
	redefining the corresponding coordinate, and the relative sign then carries no physical information.
	Throughout, we choose the orientation of the coordinate such that $g_1>0$
	and write $g_2=s|g_2|$.
	
	In the frame rotating at $\omega_L$, we describe each subensemble by its collective lowering operator $\hat S^-_\mu=\sum_{j=1}^{N_\mu}\hat\sigma^-_{\mu j}$. Since all emitters of a subensemble couple identically to the cavity and to the vibration, Eq.~\eqref{eq:Hlab} keeps each subensemble in its symmetric Dicke subspace~\cite{1954PR99,2019AQT1800043}, and we map $\hat S^-_\mu$ onto a bosonic mode through the Holstein--Primakoff transformation~\cite{1940PR1098,2010RMP1041} $\hat S^-_\mu=\sqrt{N_\mu-\hat n_\mu}\,\hat c_\mu$, with $\hat n_\mu=\hat c_\mu^\dagger\hat c_\mu$. For the low excitation numbers that enter the photon correlations, $n_\mu\ll N_\mu$, the transformation reduces to $\hat S^-_\mu\simeq\sqrt{N_\mu}\,\hat c_\mu$. The two subensembles then couple to the cavity with the collectively enhanced strengths $G_1=G\sqrt{p}$ and $G_2=G\sqrt{1-p}$, where $G=g_a\sqrt{N_1+N_2}$ and the relative abundance $p=N_1/(N_1+N_2)$ sets their ratio $G_2/G_1=\sqrt{(1-p)/p}$. The Hamiltonian in the rotating frame becomes
	\begin{equation}
		\hat H_{\rm col}=
		\Delta_a\hat a^\dagger\hat a
		+\sum_{\mu=1}^{2}\Delta_\mu\hat n_\mu
		+\omega_v\hat v^\dagger\hat v
		+\sum_{\mu=1}^{2}g_\mu\hat n_\mu(\hat v^\dagger+\hat v)
		+\sum_{\mu=1}^{2}G_\mu(\hat a^\dagger\hat c_\mu+\hat c_\mu^\dagger\hat a)
		+\Omega(\hat a^\dagger+\hat a),
		\label{eq:Hcol}
	\end{equation}
	where $\Delta_a=\omega_a-\omega_L$ and $\Delta_\mu=\omega_\mu-\omega_L$ denote the cavity and emitter detunings from the driving frequency, respectively.
	
	To identify the effective interactions mediated by the shared
	vibrational coordinate, we apply the joint polaron transformation~\cite{2002PRB235311,2016JPCM103002}
	$\hat U_{\rm P}=\exp[\sum_{\mu}\lambda_\mu\hat n_\mu(\hat v^\dagger-\hat v)]$,
	with $\lambda_\mu=g_\mu/\omega_v$. This transformation displaces the vibrational mode according to the excitations present,
	$\hat v\to\hat v-\sum_\mu\lambda_\mu\hat n_\mu$. The transformed
	Hamiltonian $\hat H_{\rm P}=\hat U_{\rm P}\hat H_{\rm col}\hat U_{\rm P}^\dagger$ reads
	\begin{align}
		\hat H_{\rm P}={}&
		\Delta_a\hat a^\dagger\hat a
		+\omega_v\hat v^\dagger\hat v
		+\sum_{\mu}\Delta_\mu\hat n_\mu
		-\chi\hat Q^2
		\nonumber\\
		&+\sum_{\mu}G_\mu
		\left(\hat a^\dagger\hat c_\mu\hat B_\mu
		+\hat c_\mu^\dagger\hat B_\mu^\dagger\hat a\right)
		+\Omega(\hat a^\dagger+\hat a),
		\label{eq:HP}
	\end{align}
	where $\hat B_\mu=\exp[-\lambda_\mu(\hat v^\dagger-\hat v)]$
	displaces the vibration whenever an excitation is exchanged
	between the cavity and subensemble $\mu$,
	$\chi=g_1^2/\omega_v$ sets the polaron shift $-\chi$ of a single
	excitation of subensemble 1, and
	$\hat Q=\hat n_1+(g_2/g_1)\hat n_2$.
	The term $-\chi\hat Q^2$ describes the energy shift
	associated with the net excitation-induced displacement
	of the shared vibrational mode.
	This displacement energy contains the polaron shifts
	$-g_\mu^2/\omega_v$ of single excitations, the intra-subensemble nonlinearities
	$U_\mu=-2g_\mu^2/\omega_v$, and the cross interaction
	$U_{12}=-2g_1g_2/\omega_v$.
	Only the cross interaction depends on $s$. The relative-displacement configuration therefore enters through the energies of states in which both subensembles are excited.
	As a reference, we compare the shared vibration with two independent vibrations, in which each subensemble couples with the same strength $g_\mu$ to its own vibrational mode of frequency $\omega_v$. The polaron transformation then replaces $-\chi\hat Q^2$ by $-\chi[\hat n_1^2+(g_2/g_1)^2\hat n_2^2]$, which contains the same polaron shifts and intra-subensemble nonlinearities $U_\mu$ but no cross interaction, $U_{12}=0$.
	
	We retain the zero-phonon-line (ZPL) contribution, in which a cavity--emitter exchange leaves the phonon number unchanged, by replacing $\hat B_\mu$ with its thermal expectation value $F_\mu=\langle\hat B_\mu\rangle_{\rm th}=\exp[-(2\bar n_v+1)\lambda_\mu^2/2]$, where $\bar n_v=[\exp(\omega_v/k_{\rm B}T)-1]^{-1}$ is the thermal occupation of the vibration at temperature $T$. This factor is the overlap of the vibrational states before and after an exchange that leaves the phonon number unchanged, averaged over the thermal distribution. Omitting the decoupled vibrational term, we obtain
	\begin{equation}
		\hat H_{\rm zpl}=
		\Delta_a\hat a^\dagger\hat a
		+\sum_{\mu}\Delta_\mu\hat n_\mu
		-\chi\hat Q^2
		+\sum_{\mu}J_\mu\left(\hat a^\dagger\hat c_\mu+\hat c_\mu^\dagger\hat a\right)
		+\Omega(\hat a^\dagger+\hat a).
		\label{eq:HZPL}
	\end{equation}
	Here $J_\mu=G_\mu F_\mu$ are the ZPL-renormalized couplings.
	This description retains the polaron shifts and the
	vibration-mediated interactions exactly. It applies when the
	phonon-assisted exchange processes, which are detuned by integer
	multiples of $\omega_v$, are weak, $(2\bar n_v+1)\lambda_\mu^2\ll1$,
	and remain off-resonant with the relevant optical transitions.
	
	We denote the Fock states of the cavity and the two collective modes by $|n_a n_1 n_2\rangle$ and the value of $\hat Q$ in such a state by $Q_{n_an_1n_2}=n_1+(g_2/g_1)n_2$. The quantity $Q_{n_an_1n_2}$ measures the displacement of the vibration in units of that produced by a single excitation of subensemble 1 and does not depend on the photon number. Each Fock state therefore has the displacement energy $-\chi Q_{n_an_1n_2}^2$.
	
	We describe the open-system dynamics within the effective
	ZPL model by including cavity loss, emitter relaxation,
	and pure dephasing. The reduced density operator $\rho$
	of the cavity and the two collective emitter modes obeys
	the Lindblad master equation
	\begin{align}
		\dot{\rho}={}&
		-i[\hat H_{\rm zpl},\rho]
		+2\kappa_a\mathcal{D}[\hat a]\rho
		+2\kappa_x\sum_{\mu=1}^{2}
		\mathcal{D}[\hat c_\mu]\rho
		+\Gamma_v\mathcal{D}[\hat Q]\rho
		+2\sum_{\mu=1}^{2}\phi_\mu
		\mathcal{D}[\hat n_\mu]\rho,
		\label{eq:ME}
	\end{align}
	where
	$\mathcal{D}[\hat O]\rho
	=\hat O\rho\hat O^\dagger
	-\{\hat O^\dagger\hat O,\rho\}/2$.
	Here, $\kappa_a$ and $\kappa_x$ are the amplitude
	decay rates of the cavity and collective emitter
	modes, respectively, with $\kappa=2\kappa_a$
	denoting the cavity linewidth. Relaxation of the individual emitters at the rate $2\kappa_x$ produces the same non-Hermitian decay $-i\kappa_x\hat n_\mu$ of each subensemble, which is the only part of the relaxation that enters the weak-driving correlations studied here.
	The rates $\phi_\mu$ describe pure dephasing caused by transition-frequency fluctuations common to all emitters within subensemble $\mu$ and statistically independent between the two subensembles; these fluctuations preserve the symmetric Dicke subspaces.
	The term $\Gamma_v\mathcal{D}[\hat Q]\rho$
	describes correlated phase fluctuations induced
	by the damped shared vibration.
	For a high-frequency vibration weakly damped by
	a thermal Markovian reservoir, the thermal fluctuations of the vibration shift the energy of each Fock state in proportion to its displacement $Q_{n_an_1n_2}$. When the cavity--emitter
	dynamics is slow compared with $\omega_v$, these shifts
	act as white noise with the dephasing rate
	$\Gamma_v=2g_1^2\gamma_v(2\bar n_v+1)/
	(\omega_v^2+\gamma_v^2)$~\cite{2010RMP1155},
	where $\gamma_v$ is the vibrational amplitude
	decay rate. The same operator $\hat Q$ thus
	governs both the coherent energy shift
	$-\chi\hat Q^2$ and the vibration-induced
	correlated dephasing.
	
	The steady-state photon statistics are characterized
	by the normalized equal-time correlation functions ($n=2,3$)
	\begin{equation}
		g^{(n)}(0)=
		\frac{\langle\hat a^{\dagger n}\hat a^n\rangle}
		{\langle\hat a^\dagger\hat a\rangle^{n}},
		\label{eq:gn}
	\end{equation}
	where $\langle\cdot\rangle$ denotes the steady-state average.
	In the weak-driving regime, cavity loss and emitter relaxation remove excitations. Their quantum jumps return the system to lower excitation manifolds and
	contribute only beyond leading order in the drive.
	In the absence of pure dephasing
	($\Gamma_v=\phi_\mu=0$), the leading-order photon
	correlations can therefore be obtained from the
	coherent amplitudes of the one-, two-, and three-excitation states~\cite{1991OC73,2011PRA021802}. These amplitudes are governed by the non-Hermitian Hamiltonian
	$\hat H_{\rm zpl}-i\kappa_a\hat a^\dagger\hat a
	-i\kappa_x\sum_\mu\hat c_\mu^\dagger\hat c_\mu$.
	By contrast, dephasing jumps preserve the excitation
	number and contribute within each manifold at the
	same order as the coherent dynamics. With finite
	dephasing, the correlations thus follow from
	the excitation-resolved steady-state density matrix
	of Eq.~\eqref{eq:ME}, including the dephasing jumps.
	Two-photon blockade (2PB) is characterized by
	$g^{(2)}(0)>1$ and $g^{(3)}(0)<1$~\cite{2019PRA053857}.

	\section{Configuration dependence across excitation manifolds}
	\label{sec:correlations}
	
	We first examine how the relative-displacement configuration enters the response of successive excitation manifolds. Throughout this section, we use the ZPL Hamiltonian of Eq.~\eqref{eq:HZPL} with cavity and emitter decay, focus on $\Gamma_v=\phi_1=\phi_2=0$, and take equal bare emitter detunings, $\Delta_1=\Delta_2=\Delta_x$. In the weak-driving limit, the steady state is described to leading order by
	$|\psi_s\rangle\simeq
	\sum C^{(s)}_{n_an_1n_2}|n_an_1n_2\rangle$,
	with $C^{(s)}_{000}\simeq1$. The amplitudes $C^{(s)}_{n_an_1n_2}$ follow from the non-Hermitian Hamiltonian $\hat H_{\rm zpl}-i\kappa_a\hat a^\dagger\hat a-i\kappa_x\sum_\mu\hat c_\mu^\dagger\hat c_\mu$ [Eq.~\eqref{eq:A_recursion}]. Since the cavity drive connects successive excitation manifolds, the amplitudes in the $m$-excitation manifold are of order $\Omega^m$. The configuration enters only through $U_{12}$, which shifts the states in which both subensembles are excited [Eq.~\eqref{eq:HP}]. The one-excitation manifold is therefore independent of $s$,  and the lowest configuration-dependent state is the mixed state $|011\rangle$, with one excitation in each subensemble.
	
	This hierarchy determines which observables can reveal the configuration. To leading order, the mean cavity population is $\langle\hat a^\dagger\hat a\rangle\simeq\Omega^2/|\Lambda_1|^2$ for both configurations. Here $\Lambda_1=\alpha-J_1^2/\beta_1-J_2^2/\beta_2$ is the complex cavity detuning dressed by exchange with the two emitter modes, with the complex detunings $\alpha=\Delta_a-i\kappa_a$ of the cavity and $\beta_\mu=\Delta_x-g_\mu^2/\omega_v-i\kappa_x$ of the polaron-shifted emitter modes. Since $\Lambda_1$ contains no $U_{12}$, the configuration affects the population only through higher-order corrections in the drive. The equal-time two-photon correlation instead compares the two-photon amplitude with the square of the one-photon amplitude, and the common drive scaling cancels. We find
	\begin{equation}
		g_s^{(2)}(0)=|\Lambda_1|^2\,\frac{|1+\mathcal R_{2,s}|^2}{|\alpha-\Sigma_{2,s}|^2}.
		\label{eq:g2_weak}
	\end{equation}
	Here $\alpha-\Sigma_{2,s}$ is the complex energy per photon of the two-photon state $|200\rangle$, dressed by virtual exchange with the other two-excitation states. These are $|110\rangle$ and $|101\rangle$, with one photon and one emitter excitation, and $|020\rangle$, $|011\rangle$, and $|002\rangle$, with two emitter excitations. The ratio $\mathcal R_{2,s}$ is the amplitude of the indirect pathway relative to the direct pathway $|000\rangle\to|100\rangle\to|200\rangle$. In the indirect pathway, the drive adds a photon to a state that already contains an emitter excitation, and this excitation subsequently returns to the cavity. The factor $1+\mathcal R_{2,s}$ thus describes the interference between the two pathways. Both $\Sigma_{2,s}$ and $\mathcal R_{2,s}$ [Eq.~\eqref{eq:A_SigmaR}] depend on $s$ only through the energy of $|011\rangle$, which is coupled to $|200\rangle$ via $|110\rangle$ and $|101\rangle$. The normalized two-photon correlation therefore depends in general on the configuration already in the weak-driving limit.
	
	Because the mixed state $|011\rangle$ contains no photon, the drive cannot populate it directly. The mixed state is reached only when the photon in $|110\rangle$ or $|101\rangle$ is absorbed by the other subensemble. Its amplitude is therefore the coherent sum of two exchange pathways,
	\begin{equation}
		C_{011}^{(s)}=-\frac{J_2C_{110}^{(s)}+J_1C_{101}^{(s)}}{\beta_1+\beta_2+U_{12}},
		\label{eq:C011}
	\end{equation}
	where $\beta_1+\beta_2+U_{12}$ is the complex energy of the mixed state. When the two pathways interfere destructively, $J_2C_{110}+J_1C_{101}=0$, the mixed state remains empty, and the two-excitation response is screened from the finite cross interaction $U_{12}$. With $|011\rangle$ empty, the remaining two-excitation amplitudes no longer involve $U_{12}$. The screening condition [Eq.~\eqref{eq:A_screen}] is therefore the same for both configurations. Screening requires the two pathways to arrive with equal magnitudes and opposite phases. This is possible for unequal displacement magnitudes, $g_1^2\neq g_2^2$: for equal bare emitter detunings, the states $|110\rangle$ and $|101\rangle$, through which the two pathways pass, then carry different polaron shifts, $\beta_1-\beta_2=-(g_1^2-g_2^2)/\omega_v$ [Eq.~\eqref{eq:A_ratio}]. For given couplings $g_\mu$ to the vibration and emitter--cavity detuning $\Delta_x-\Delta_a$, the balance of magnitudes and phases fixes $J_1$ and $J_2$ at each cavity detuning $\Delta_a$ [Eq.~\eqref{eq:A_Jscreen}], and hence $G$ and $p$ through the ZPL-renormalized couplings. The solutions with positive $J_1^2$ and $J_2^2$ form the screened branch, which we label by $G/\kappa$.
	
\begin{figure}
		\centering
		\includegraphics[height=0.43\linewidth]{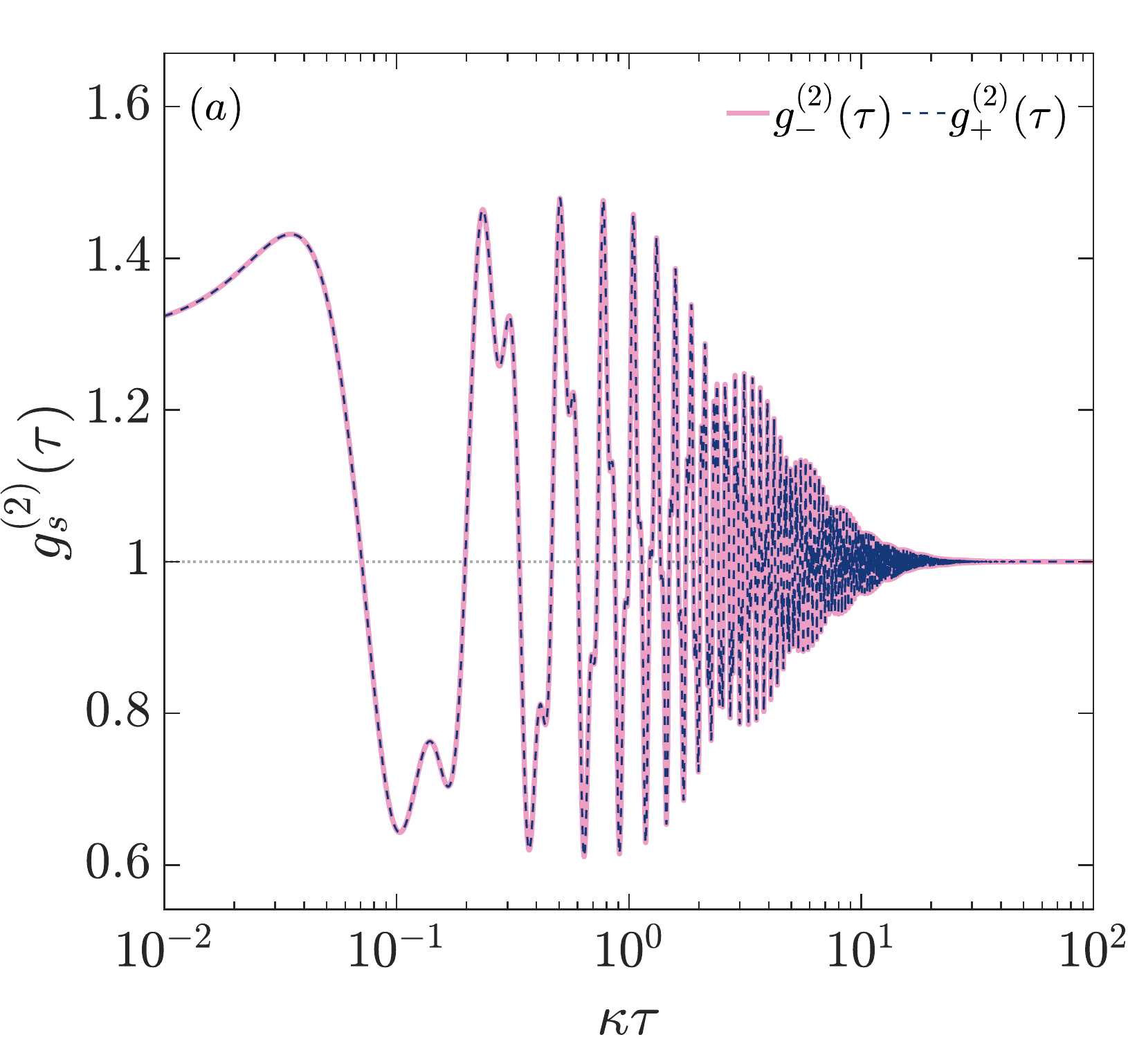}
		\hfill
		\includegraphics[height=0.43\linewidth]{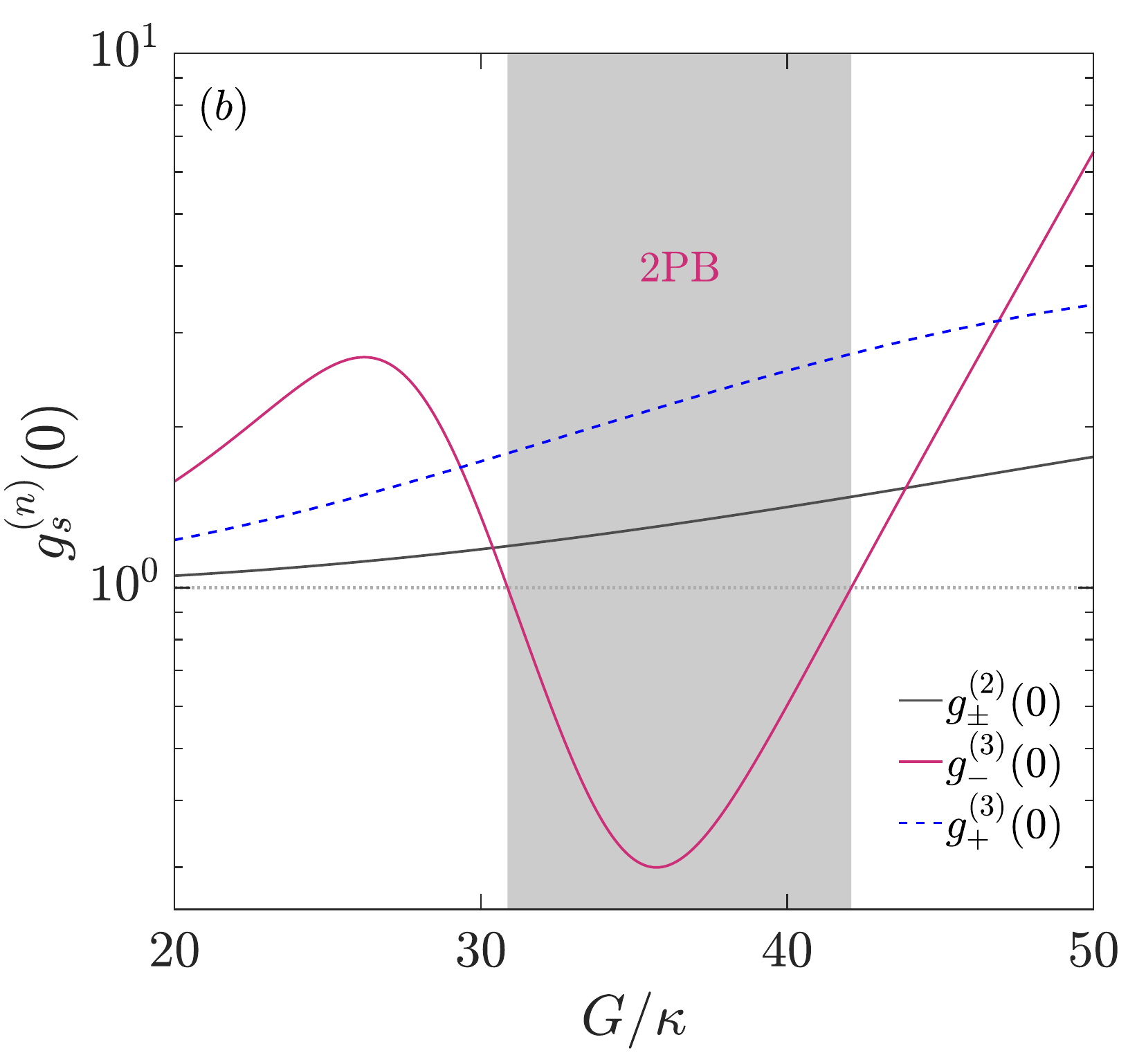}
		\caption{
			(a) Delayed correlation $g_s^{(2)}(\tau)$ for $s=\pm1$ at $\Delta_a=\chi$ and $\Delta_x=2.8\chi$, corresponding to $G/\kappa\simeq36$ in (b). (b) Equal-time correlations $g_\pm^{(2)}(0)$ and $g_\pm^{(3)}(0)$ versus $G/\kappa$ along the screened branch with $\Delta_x-\Delta_a=1.8\chi$, on which $\Delta_a$ changes with $G/\kappa$; in the shaded interval, $g_-^{(3)}(0)<1<g_+^{(3)}(0)$. In both panels, $J_1$ and $J_2$ satisfy the screening condition of Eq.~\eqref{eq:A_Jscreen}, which keeps $|011\rangle$ empty, and all curves follow from the weak-driving amplitudes of Eq.~\eqref{eq:A_recursion}; in (a), the conditional state of Eq.~\eqref{eq:A_cond} evolves in the one-excitation manifold of Eq.~\eqref{eq:A_H1}. Other parameters: $|g_2|=2g_1$, $\omega_v=40\chi$, $\bar n_v=0$, $\chi/\kappa=30$, and $\kappa_x/\kappa=0.02$.
		}
		\label{fig:screening}
	\end{figure}

	On the screened branch, all leading-order two-excitation amplitudes are identical for $s=\pm1$, and hence $g_+^{(2)}(0)=g_-^{(2)}(0)$. The screening extends to the delayed correlation $g_s^{(2)}(\tau)=\langle\hat a^\dagger(0)\hat a^\dagger(\tau)\hat a(\tau)\hat a(0)\rangle_s/\langle\hat a^\dagger\hat a\rangle_s^2$, which probes the evolution of the system after a photon detection. Since this photon-conditioned dynamics is governed by the $s$-independent one-excitation manifold, the delayed correlations of the two configurations share the same oscillation frequencies and decay rates and can differ only in their amplitudes. On the screened branch, a photon detection leaves the system in the same conditional state for both configurations [Eq.~\eqref{eq:A_cond}], and these amplitudes coincide as well. Figure~\ref{fig:screening}(a) shows that $g_+^{(2)}(\tau)$ and $g_-^{(2)}(\tau)$ coincide at all delays. The two-excitation sector is thus insensitive to the relative-displacement configuration.
	
	The three-excitation response remains configuration dependent on the screened branch, because the asymmetry $g_1^2\neq g_2^2$ that permits the screening prevents its extension to this manifold. In this manifold, the configuration enters through the three mixed states $|111\rangle$, $|021\rangle$, and $|012\rangle$, in which both subensembles are excited. The photonless mixed states $|021\rangle$ and $|012\rangle$ also receive contributions that start from $|020\rangle$ and $|002\rangle$, the states with two excitations in one subensemble: the drive adds a photon to these states, and the photon is then absorbed by the other subensemble, $|020\rangle\to|120\rangle\to|021\rangle$ and $|002\rangle\to|102\rangle\to|012\rangle$. The states $|020\rangle$ and $|002\rangle$ displace the vibration by $Q_{020}=2$ and $Q_{002}=2g_2/g_1$, and their displacement energies differ by
	\begin{equation}
		\chi\left|Q_{020}^2-Q_{002}^2\right|
		=\frac{4|g_1^2-g_2^2|}{\omega_v}.
		\label{eq:mismatch}
	\end{equation}
	Complete screening of the three-excitation response would require $|111\rangle$, $|021\rangle$, and $|012\rangle$ to remain empty, in analogy with $|011\rangle$. Exchange from $|210\rangle$ and $|201\rangle$ must then cancel the drive of $|120\rangle$ and $|102\rangle$ from $|020\rangle$ and $|002\rangle$, and the two exchange contributions to $|111\rangle$, from $|210\rangle$ and from $|201\rangle$, must cancel each other. With the screened two-excitation amplitudes, both cancellations hold only if $|020\rangle$ and $|002\rangle$ are degenerate [Eqs.~\eqref{eq:A_H2} and \eqref{eq:A_feed3}], which Eq.~\eqref{eq:mismatch} excludes for $g_1^2\neq g_2^2$. The leading-order three-excitation amplitudes of the two configurations therefore differ, whereas their two-excitation responses coincide.
	
	Figure~\ref{fig:screening}(b) shows how this difference appears in the photon statistics along the screened branch. Both configurations exhibit the same two-photon bunching, $g_+^{(2)}(0)=g_-^{(2)}(0)>1$. Within the shaded interval of $G/\kappa$, $g_-^{(3)}(0)<1<g_+^{(3)}(0)$, and only the $s=-1$ configuration exhibits 2PB.
	
	This selectivity arises from quantum interference between two excitation pathways of the three-photon state: a direct pathway and an indirect pathway through the mixed state $|111\rangle$. Apart from $|300\rangle$ itself, $|111\rangle$ is the only three-excitation state that couples the states reached from $|300\rangle$ by exciting subensemble 1, $|210\rangle$ and $|120\rangle$, to those reached by exciting subensemble 2, $|201\rangle$ and $|102\rangle$. Exact elimination of these two pairs and of the photonless states reduces the three-photon response to two coupled channels, $|300\rangle$ and $|111\rangle$ [Eq.~\eqref{eq:A_twochannel}], and yields
	\begin{equation}
		g_s^{(3)}(0)=g_{{\rm bg},s}^{(3)}\left|\frac{\mathcal E_{111}-\mathcal Z}{\mathcal E_{111}-\mathcal P}\right|^2.
		\label{eq:g3_twochannel}
	\end{equation}
	Here $g_{{\rm bg},s}^{(3)}$ is the three-photon correlation in the absence of the indirect pathway, and $\mathcal E_{111}$ is the complex energy of $|111\rangle$ dressed by virtual exchange with the other three-excitation states. The indirect pathway cancels the direct one when $\mathcal E_{111}$ reaches the interference zero $\mathcal Z$. The denominator contains $\mathcal E_{111}-\mathcal P$, the energy of $|111\rangle$ further dressed by its coupling to $|300\rangle$ [Eq.~\eqref{eq:A_g3}]. On the screened branch, the drive does not populate $|111\rangle$ from $|011\rangle$, since $C_{011}=0$. The bare energy of $|111\rangle$ differs between the two configurations by $4g_1|g_2|/\omega_v$ [Eq.~\eqref{eq:A_H3diff}], and its dressed energy $\mathcal E_{111}$ acquires further configuration-dependent shifts through virtual exchange with the photonless states $|021\rangle$ and $|012\rangle$. The background $g_{{\rm bg},s}^{(3)}$ and the interference zero $\mathcal Z$ depend on the configuration only through these photonless states, and their configuration dependence is much weaker than that of $\mathcal E_{111}$. Within the shaded interval of Fig.~\ref{fig:screening}(b), the indirect pathway interferes destructively with the direct one for $s=-1$ and suppresses $g_-^{(3)}(0)$ below unity. The cancellation is strongest near the point of Fig.~\ref{fig:screening}(a), where $\mathcal E_{111}$ lies close to $\mathcal Z$. For $s=+1$, the configuration-dependent shift moves $\mathcal E_{111}$ far from $\mathcal Z$, and the destructive interference is lost. Replacing only the bare energy of $|111\rangle$ in the $s=+1$ configuration by its $s=-1$ value restores the suppression (Appendix~\ref{app:three}). This identifies $|111\rangle$ as the state that predominantly mediates the configuration selectivity. Together with Fig.~\ref{fig:screening}(a), these results establish a configuration dependence that is selective in the excitation order: the leading-order two-photon correlations of the cavity field coincide at all delays, whereas the three-photon statistics distinguish the two configurations through the interference mediated by $|111\rangle$.
	
	\section{Displacement-compensated three-photon resonance}
	\label{sec:threephoton}
	
	Section~\ref{sec:correlations} showed that coherent interference can make the leading-order two-excitation responses of the two configurations identical, whereas their three-photon statistics differ. We now show that, in the dispersive regime $J\ll\chi$, where $J$ denotes the scale of the exchange couplings $J_1$ and $J_2$, the shared vibration enables a resonant conversion of photon triplets into emitter excitations. For $s=-1$, one of the photonless mixed states through which the configuration enters the three-excitation response, $|021\rangle$ with two excitations in subensemble 1 and one in subensemble 2, becomes resonant with the three-photon state $|300\rangle$. We refer to $|021\rangle$ as the composite. The coherent conversion of three photons into the composite suppresses $g^{(3)}(0)$, while $g^{(2)}(0)$ remains close to unity. The shared vibration provides both the resonance and the coupling that this conversion requires. For $s=+1$, the composite lies far from resonance at the same detunings. For two independent vibrations, the composite can be tuned into resonance, but its conversion coupling vanishes to leading order.
	\subsection{Resonance from displacement neutrality}
	\label{sec:resonance}
\begin{figure*}
		\centering
		\includegraphics[width=\linewidth]{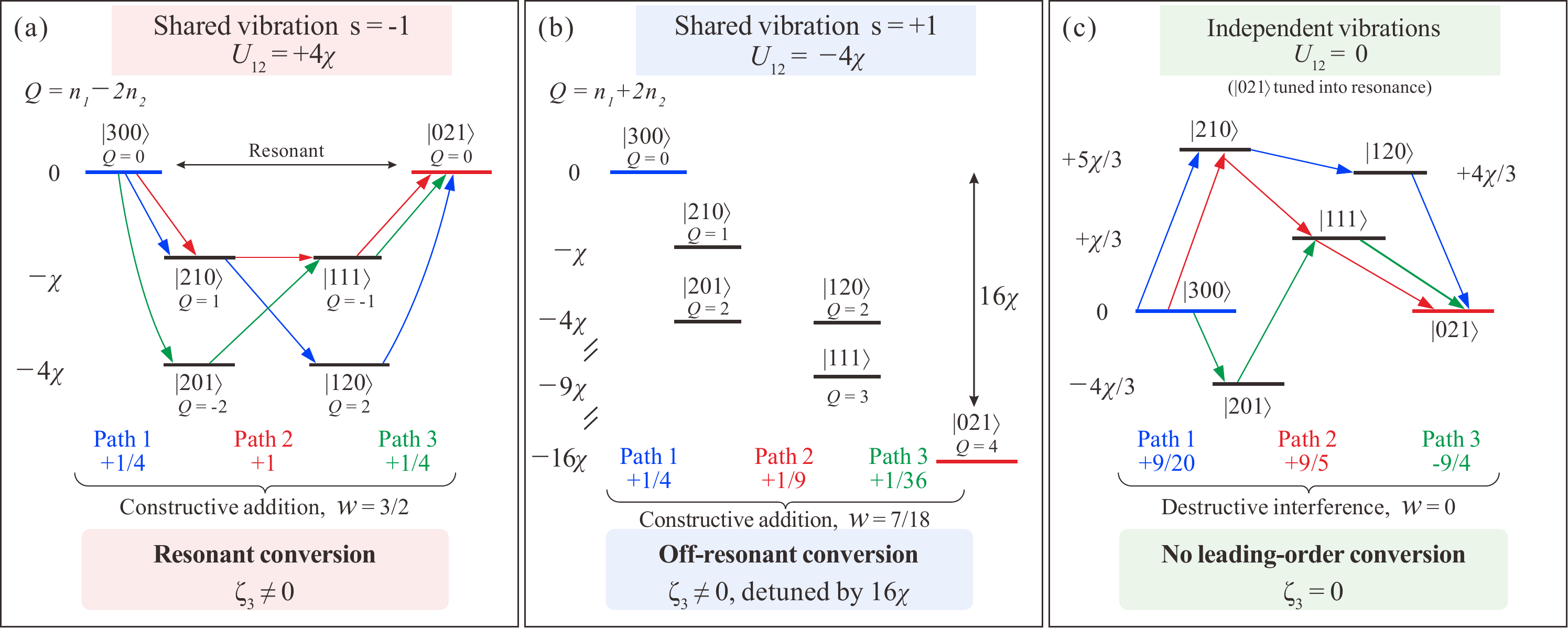}
		\caption{Level schemes of the three-excitation manifold and conversion paths from $|300\rangle$ to $|021\rangle$. The three-photon state $|300\rangle$ and the composite $|021\rangle$ form the resonant pair. The intermediate states $|210\rangle$, $|201\rangle$, $|120\rangle$, and $|111\rangle$ are reached from the resonant pair by a single cavity--emitter exchange and are populated only virtually. (a) Shared vibration with $s=-1$: all intermediate states lie below the resonant pair, and the three paths add. (b) Shared vibration with $s=+1$: the three paths still add, but $|021\rangle$ lies $16\chi$ below $|300\rangle$. (c) Independent vibrations ($U_{12}=0$), with the emitters detuned so that $|021\rangle$ and $|300\rangle$ are degenerate: path~3 changes sign, and the three paths cancel. Paths 1, 2, and 3 denote $|300\rangle\to|210\rangle\to|120\rangle\to|021\rangle$, $|300\rangle\to|210\rangle\to|111\rangle\to|021\rangle$, and $|300\rangle\to|201\rangle\to|111\rangle\to|021\rangle$; the numbers give the contributions $\chi^2/(d_{210}d_{120})$, $\chi^2/(d_{210}d_{111})$, and $\chi^2/(d_{201}d_{111})$ of the three paths to their net weight $\mathcal W$ [Eq.~\eqref{eq:B_W}], with $d_\nu$ the energies of the intermediate states measured from the resonant pair. Level spacings are not to scale.}
		\label{fig:paths}
	\end{figure*}

	We take $|g_2|=2g_1$, as in Fig.~\ref{fig:screening}, so that $\hat Q=\hat n_1+2s\hat n_2$. This is the smallest of the integer ratios discussed at the end of this section. For $s=-1$, one excitation of subensemble 2 displaces the vibration twice as far as one excitation of subensemble 1, and in the opposite direction. The composite $|021\rangle$ therefore leaves the vibration undisplaced, $Q_{021}=0$, and its displacement energy $-\chi Q_{021}^2$ vanishes, as does that of the three-photon state $|300\rangle$. For $s=-1$, $Q=0$ requires $n_1=2n_2$; every undisplaced state with emitter excitations contains at least three of them, and $|021\rangle$ is the only such state in the three-excitation manifold. When the emitters are resonant with the cavity, $|021\rangle$ and $|300\rangle$ are thus degenerate [Fig.~\ref{fig:paths}(a)]. For $s=+1$, the two displacements add, $Q_{021}=4$, and at the same detunings $|021\rangle$ lies $16\chi$ below $|300\rangle$ [Fig.~\ref{fig:paths}(b)]. For two independent vibrations, each subensemble displaces its own vibration. The composite then retains the displacement energy $-8\chi$, and its resonance with $|300\rangle$ requires the emitter--cavity detuning $\Delta_x-\Delta_a\simeq8\chi/3$ [Fig.~\ref{fig:paths}(c) and Eq.~\eqref{eq:B_detunings_ind}].
	
	Virtual cavity--emitter exchange, in which the system passes briefly through an off-resonant state and returns, shifts both members of the resonant pair, $|300\rangle$ and $|021\rangle$. For $s=-1$, the resonance requires these shifts to be compensated. A single exchange connects $|300\rangle$ to $|210\rangle$ and $|201\rangle$, and $|021\rangle$ to $|111\rangle$ and $|120\rangle$ [Fig.~\ref{fig:paths}(a)]. These intermediate states lie below the resonant pair, $|210\rangle$ and $|111\rangle$ by $\chi$ and $|201\rangle$ and $|120\rangle$ by $4\chi$, and virtual exchange with them shifts both members of the pair upward. To second order in $J_1$ and $J_2$, the one-photon level, and hence each photon in $|300\rangle$, is raised by $J_1^2/\chi+J_2^2/(4\chi)$ through virtual excitation of the two emitter modes [Eq.~\eqref{eq:B_E100}]. The composite is raised by $2J_1^2/\chi+J_2^2/(4\chi)$ through $|111\rangle$ and $|120\rangle$ [Eq.~\eqref{eq:B_E021}]. These Lamb shifts are compensated to leading order by
	\begin{equation}
		\Delta_a\simeq-\frac{J_1^2+J_2^2/4}{\chi},\qquad
		\Delta_x\simeq-\frac{2J_1^2+J_2^2/4}{3\chi}.
		\label{eq:resonance_compensation}
	\end{equation}
	Here $\Delta_a$ cancels the Lamb shift of the photon and thereby keeps the drive resonant with the dressed cavity. The detuning $3\Delta_x$ cancels the Lamb shift of the composite and thereby keeps $|021\rangle$ degenerate with three dressed photons. The required emitter--cavity detuning, $\Delta_x-\Delta_a\simeq(J_1^2+J_2^2/2)/(3\chi)$, is thus of order $J^2/\chi$, much smaller than the detuning $8\chi/3$ required for two independent vibrations.
	
	\begin{figure}
		\centering
		\includegraphics[width=0.49\linewidth]{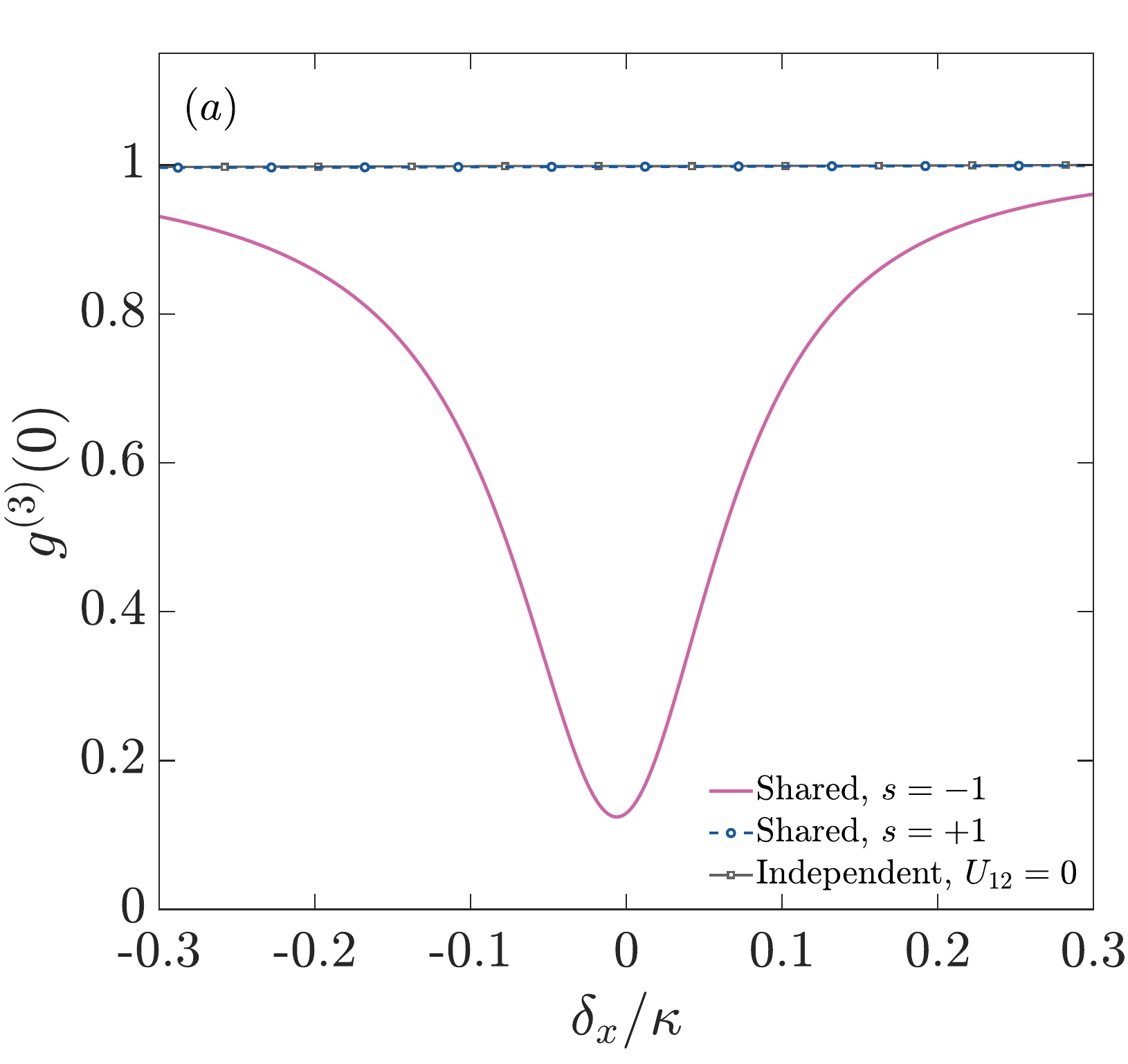}
		\hfill
		\includegraphics[width=0.49\linewidth]{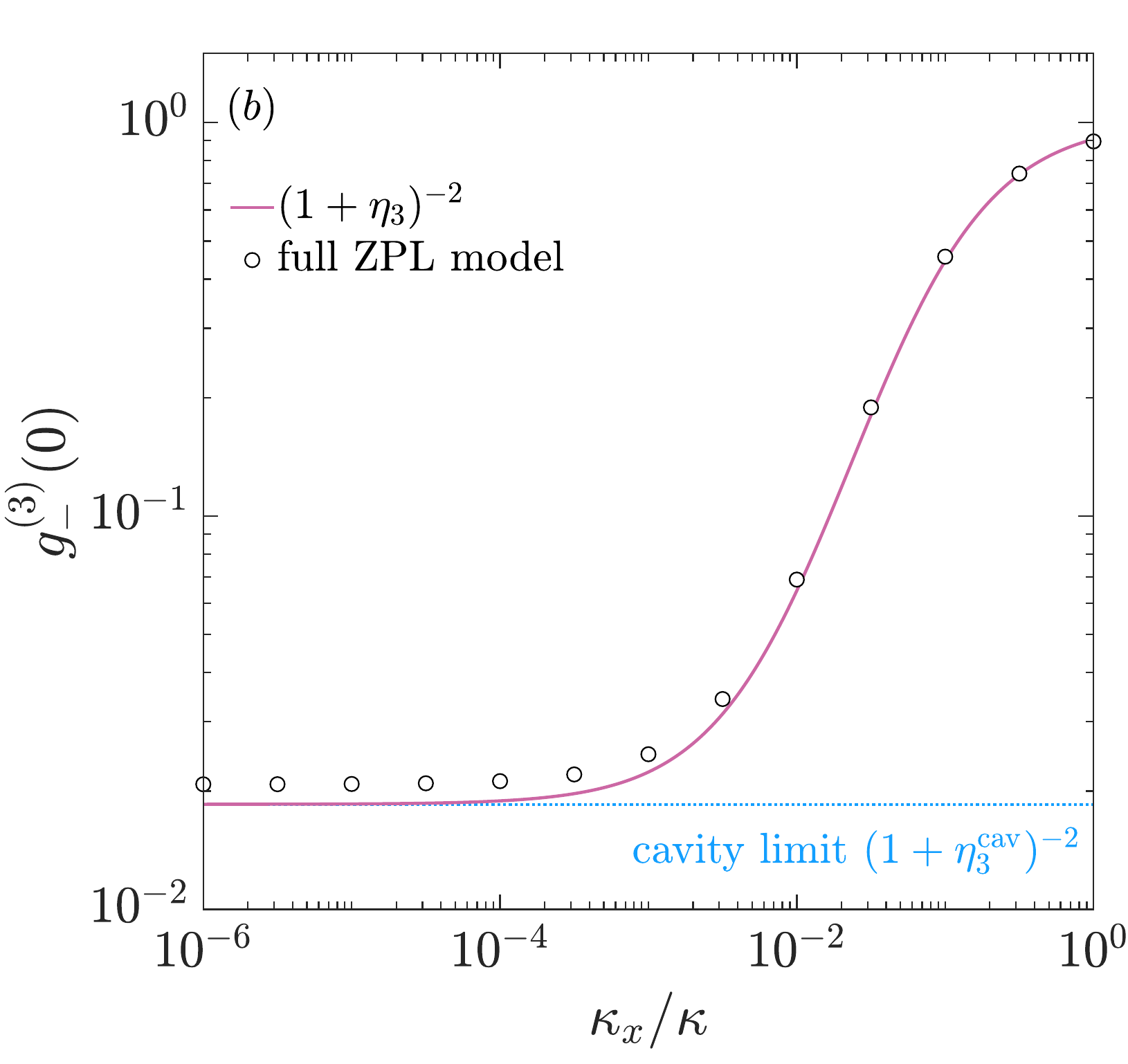}
		\caption{
			(a) Three-photon correlation $g^{(3)}(0)$ versus the deviation $\delta_x$ of the emitter detuning from its compensated value, at the compensated cavity detuning, for the shared vibration with $s=\pm1$ ($U_{12}=-4s\chi$), with the detunings of Eq.~\eqref{eq:resonance_compensation}, and for two independent vibrations ($U_{12}=0$), with the detunings at which $|021\rangle$ and $|300\rangle$ are degenerate [Eq.~\eqref{eq:B_detunings_ind}]. (b) $g_-^{(3)}(0)$ at the resonance of $|021\rangle$ and $|300\rangle$ versus the emitter decay rate $\kappa_x$. The curves in (a) and the circles in (b) follow from the weak-driving amplitudes of Eq.~\eqref{eq:A_recursion}; the solid line in (b) follows from Eq.~\eqref{eq:control_law} and the dotted line from its cavity limit, Eq.~\eqref{eq:cavity_limit}, both with the dressed parameters of Eq.~\eqref{eq:B_Heff}. Parameters: $J_1=J_2=0.15\chi$; other parameters as in Fig.~\ref{fig:screening}.
		}
		\label{fig:threephoton}
	\end{figure}
	
	\subsection{Conversion through interfering exchange paths}
	\label{sec:conversion}
	Figure~\ref{fig:threephoton}(a) shows $g^{(3)}(0)$ as a function of the deviation $\delta_x$ of the emitter detuning from its compensated value, at the compensated cavity detuning. For both configurations of the shared vibration, the compensated detunings are those of Eq.~\eqref{eq:resonance_compensation}, at which the $s=-1$ composite is resonant with $|300\rangle$. For two independent vibrations, the compensated detunings of Eq.~\eqref{eq:B_detunings_ind} bring the composite into resonance with $|300\rangle$. For $s=-1$, $g^{(3)}(0)$ develops a pronounced dip at $\delta_x=0$, where $|021\rangle$ is resonant with $|300\rangle$. No such dip appears for $s=+1$, whose composite lies far off resonance, or for two independent vibrations, whose composite is resonant with $|300\rangle$ at $\delta_x=0$. The dip therefore requires a coupling of the resonant pair in addition to the resonance.
	
	This coupling arises from three successive cavity--emitter exchanges, which convert three photons into two excitations of subensemble 1 and one of subensemble 2. Since the subensemble-2 excitation can be created first, second, or last, the conversion proceeds along the three paths of Fig.~\ref{fig:paths}. Eliminating their intermediate states, which are detuned by energies of order $\chi$ and populated only virtually, gives the effective conversion Hamiltonian
	\begin{equation}
		H_{\rm conv}=\zeta_3\left(\hat c_1^{\dagger2}\hat c_2^\dagger\hat a^3+{\rm H.c.}\right),
		\label{eq:Hconv}
	\end{equation}
	which removes three photons from the cavity and creates the composite, or the reverse. At resonance, photon triplets leak through the conversion into the decaying composite, which acts as a resonantly enhanced loss of the three-photon state: its rate is set by the squared conversion strength divided by the composite decay rate [Eq.~\eqref{eq:control_law}], so that a narrower composite gives a stronger loss, and this loss suppresses $g^{(3)}(0)$. The conversion strength $\zeta_3$ results from the interference of the three paths. The cross interaction $U_{12}$ controls this interference, since the energy cost of creating the subensemble-2 excitation changes by $U_{12}$ with each excitation already present in subensemble 1. At the bare resonance, where $|021\rangle$ and $|300\rangle$ are degenerate before the Lamb shifts are included, summing the three paths gives [Eqs.~\eqref{eq:B_W} and \eqref{eq:B_WU12}]
	\begin{equation}
		\zeta_3\simeq J_1^2J_2\,\frac{U_{12}\left(U_1+2U_{12}\right)}{d_{210}d_{120}d_{111}d_{201}},
		\label{eq:V3_shared}
	\end{equation}
	where $U_1+2U_{12}$ is the interaction energy of the composite and $d_\nu$ are the energies of the intermediate states measured from the resonant pair. For $s=-1$, $U_{12}=4\chi$, the three paths add, and their net weight $\mathcal W=\zeta_3\chi^2/(J_1^2J_2)$ equals $3/2$ [Fig.~\ref{fig:paths}(a)]; hence $\zeta_3\simeq3J_1^2J_2/(2\chi^2)$. For $s=+1$, $U_{12}=-4\chi$, and at the same detunings the three paths also add, with $\mathcal W=7/18$ [Eq.~\eqref{eq:B_W} and Fig.~\ref{fig:paths}(b)], but the resonance of the composite with $|300\rangle$ lies at the emitter--cavity detuning $\Delta_x-\Delta_a\simeq16\chi/3$ [Eq.~\eqref{eq:B_dnu}], far from the compensated detunings of Eq.~\eqref{eq:resonance_compensation}. For two independent vibrations, $U_{12}=0$, and the three paths cancel, $\mathcal W=0$ [Fig.~\ref{fig:paths}(c)]. The composite therefore combines a resonance with $|300\rangle$ and a finite conversion strength only for $s=-1$, and the dip of Fig.~\ref{fig:threephoton}(a) appears for this configuration alone. For $s=-1$, the cross interaction $U_{12}$, whose sign is set by the configuration, thus both cancels the displacement energy of the composite and generates its coupling to the three-photon state.
	
	The same virtual exchanges leave $g^{(2)}(0)$ close to unity in all three cases. A single virtual emitter excitation shifts the photonic levels in proportion to the photon number and leaves the photon ladder harmonic. An anharmonic two-photon shift requires two virtual emitter excitations that are present at the same time and interact, and hence four exchanges. This residual two-photon shift scales as $J^4/\chi^3$ [Eq.~\eqref{eq:B_Kerr}], one order in $J/\chi$ beyond $\zeta_3$. Without the interactions $U_1$, $U_2$, and $U_{12}$, the driven cavity and emitter modes remain in a coherent state with $g^{(2)}(0)=1$. The deviation of $g^{(2)}(0)$ from unity therefore arises only from these four-exchange processes and is controlled by four powers of $J/\chi$ [Eq.~\eqref{eq:B_g2dev}]. In all three cases of Fig.~\ref{fig:threephoton}(a), $g^{(2)}(0)$ thus deviates from unity by at most about $10^{-3}$ [Fig.~\ref{fig:g2}], and the configuration dependence appears only in the three-photon statistics.
	
	\subsection{Depth of the three-photon dip}
	\label{sec:depth}
	The depth of the dip for $s=-1$ follows from the dressed effective Hamiltonian of Eq.~\eqref{eq:B_Hdressed}, in which the cavity is coupled to the composite only through the conversion. At resonance, the conversion transfers the decay of the composite to the three-photon amplitude, and Eq.~\eqref{eq:B_control} gives
	\begin{equation}
		g_-^{(3)}(0)\simeq\frac{1}{(1+\eta_3)^2}.
		\label{eq:control_law}
	\end{equation}
	Here $\eta_3\simeq4\zeta_3^2/(\kappa_a\Gamma_{021})$ is the three-photon cooperativity, which compares the squared conversion strength with the product of the cavity decay rate $\kappa_a$ and the composite decay rate $\Gamma_{021}$ [Eq.~\eqref{eq:B_eta3}]. Equation~\eqref{eq:control_law} has the same form as the suppression of the intracavity intensity by a single resonant absorber. The composite decay rate $\Gamma_{021}\simeq3\kappa_x+(\kappa_a+2\kappa_x)(2J_1^2+J_2^2/16)/\chi^2$ [Eq.~\eqref{eq:B_bT}] has two contributions: the decay $3\kappa_x$ of its three emitter excitations, and the cavity loss inherited from the intermediate states $|111\rangle$ and $|120\rangle$. Virtual exchange admixes these states, each containing one cavity photon, into the composite with the probabilities $2J_1^2/\chi^2$ and $J_2^2/(16\chi^2)$.
	
	Figure~\ref{fig:threephoton}(b) shows that the full weak-driving results follow Eq.~\eqref{eq:control_law} from broad to narrow emitter lines. For large $\kappa_x$, the composite decays mainly through its own emitter excitations, $\Gamma_{021}\simeq3\kappa_x$, and $\eta_3\simeq4\zeta_3^2/(3\kappa_a\kappa_x)\ll1$. The composite then acts as a broad and weak absorber: the damping that the conversion adds to the three-photon amplitude is small compared with the cavity loss, and $g_-^{(3)}(0)$ approaches unity, the value for the nearly harmonic dressed cavity. As $\kappa_x$ decreases, $\eta_3$ initially grows approximately as $1/\kappa_x$, and the dip deepens until the inherited cavity loss dominates $\Gamma_{021}$. For $\kappa_x\to0$, the composite decays only through this inherited cavity loss, and the dip saturates at the cavity limit [Eq.~\eqref{eq:B_eta3cav}]
	\begin{equation}
		g_-^{(3)}(0)\simeq\frac{1}{(1+\eta_3^{\rm cav})^2},\qquad
		\eta_3^{\rm cav}\simeq\frac{4\zeta_3^2\chi^2}{\kappa_a^2\left(2J_1^2+J_2^2/16\right)}.
		\label{eq:cavity_limit}
	\end{equation}
	With $\zeta_3\simeq3J_1^2J_2/(2\chi^2)$, the cavity-limited cooperativity is of order $J^4/(\chi\kappa_a)^2$. A deep dip, $\eta_3^{\rm cav}\gg1$, therefore requires $\chi/\kappa_a\gg(\chi/J)^2$. Relative to the cavity decay rate, the residual two-photon shift of Eq.~\eqref{eq:B_Kerr} is of order $(J/\chi)^2(\eta_3^{\rm cav})^{1/2}$, and Eq.~\eqref{eq:B_g2dev} gives $|g^{(2)}(0)-1|\sim(J/\chi)^4(1+\eta_3^{\rm cav})$. For $s=-1$, the deep three-photon dip therefore coexists with $g^{(2)}(0)\simeq1$ in the window $1\ll\eta_3^{\rm cav}\ll(\chi/J)^4$, which is wide in the dispersive regime.
	
	\subsection{Family of displacement-compensated conversions}
	\label{sec:family}
	The ratio $|g_2|=2g_1$ is the smallest integer ratio compatible with the screening of Sec.~\ref{sec:correlations}, which requires $g_1^2\neq g_2^2$. It produces the first of a family of displacement-compensated resonances. For an integer ratio $r=|g_2|/g_1\geq2$ and $s=-1$, $\hat Q=\hat n_1-r\hat n_2$, and the role of the composite is taken by $|0,r,1\rangle$, with $r$ excitations in subensemble 1 and one in subensemble 2. This state is the only undisplaced state with emitter excitations in the $(r+1)$-excitation manifold and is degenerate with the $(r+1)$-photon state $|r{+}1,0,0\rangle$ when the emitters are resonant with the cavity. For $s=+1$, the same state carries $Q=2r$ and lies $4r^2\chi$ below the $(r+1)$-photon state. For $s=-1$, the $r+1$ conversion paths differ only in the step at which the subensemble-2 excitation is created. All of them pass through displaced intermediate states below the resonant pair, and they add. The conversion Hamiltonian $\zeta_{r+1}(\hat c_1^{\dagger r}\hat c_2^\dagger\hat a^{r+1}+{\rm H.c.})$, which generalizes Eq.~\eqref{eq:Hconv}, then has the strength [Eq.~\eqref{eq:B_Wr_shared}]
	\begin{equation}
		\zeta_{r+1}\simeq\frac{(2r)!}{(r!)^4}\,\frac{J_1^rJ_2}{\chi^r},
		\label{eq:zeta_r}
	\end{equation}
	which reduces to $\zeta_3\simeq3J_1^2J_2/(2\chi^2)$ for $r=2$. For two independent vibrations, the subensemble-2 excitation costs the same energy at every step, and the paths sum to an amplitude proportional to the energy of $|0,r,1\rangle$ measured from the $(r+1)$-photon state, which vanishes at the bare resonance [Eq.~\eqref{eq:B_Wr_ind}]. For each integer $r$, the shared vibration with $s=-1$ thus provides both the resonance and the conversion, and displacement neutrality sets the number of converted photons. This holds within the ZPL description, $(2\bar n_v+1)r^2\chi/\omega_v\ll1$, for $r+1\ll N_\mu$, and in the dispersive regime $\sqrt{r}\,J\ll\chi$, where the bosonically enhanced exchanges remain weak compared with the displacement energies. At fixed exchange couplings, each further photon adds a factor of order $J/\chi$ to the conversion strength, and the combinatorial factor in Eq.~\eqref{eq:zeta_r} decreases with $r$. For $r=1$, the same displacement compensation makes $|011\rangle$ resonant with $|200\rangle$, and the conversion of photon pairs suppresses $g^{(2)}(0)$ itself, so that the three-photon conversion is the strongest one that leaves the two-photon statistics nearly Poissonian.
	
	Near an integer ratio, $|g_2|/g_1=r+\epsilon$ with $|\epsilon|\ll1$, the composite $|0,r,1\rangle$ retains the displacement $-\epsilon$. Its displacement energy $-\chi\epsilon^2$ is compensated by a small change of the emitter detuning, and the conversion pathways still pass through intermediate states below the resonant pair and add, with a leading-order strength that changes linearly with $\epsilon$ [Eq.~\eqref{eq:B_Wr}]. 
	\section{Displacement-selective dephasing of the three-photon resonance}
	\label{sec:dephasing}
	
	The shared vibration enters the master equation~\eqref{eq:ME} twice: its displacement generates the interaction $-\chi\hat Q^2$ that produces the three-photon resonance, and its thermal fluctuations generate the dephasing $\Gamma_v\mathcal D[\hat Q]$. A fluctuation of the vibrational coordinate shifts the energy of each state in proportion to its displacement and thus randomizes the relative phase of two states $|\nu\rangle$ and $|\nu'\rangle$ at the rate $\Gamma_v(Q_\nu-Q_{\nu'})^2/2$, which vanishes for equally displaced states [Eq.~\eqref{eq:C_damping}]. For $s=-1$, the resonant pair, the three-photon state $|300\rangle$ and the composite $|021\rangle$, consists of two undisplaced states, $Q_{300}=Q_{021}=0$. The displacement compensation that makes the pair degenerate therefore also cancels the vibrational dephasing between its two states.
	
	The vibrational noise reaches the resonant pair only through the displaced intermediate states that virtual exchange admixes into it with weights of order $(J/\chi)^2$: $|111\rangle$ and $|120\rangle$ into the composite, and $|210\rangle$ and $|201\rangle$ into the three-photon state. Scattering out of these admixtures removes the system from the resonant pair, and this vibrational leakage acts as an additional decay of both states. For the photons, the leakage renormalizes $\kappa_a$. Emitter dephasing acts on the excitation numbers $\hat n_\mu$: the two subensembles fluctuate independently, and the composite acquires the random phase of subensemble 1 twice and that of subensemble 2 once. To leading order in $J/\chi$, the composite therefore decays and loses its phase relative to the photonic states $|n00\rangle$ at the two rates [Eq.~\eqref{eq:C_rates}]
	\begin{equation}
		\Gamma_{021}\simeq3\kappa_x+(\kappa_a+2\kappa_x)\frac{2J_1^2+J_2^2/16}{\chi^2}+\frac{\Gamma_v}{2}\,\frac{2J_1^2+J_2^2/4}{\chi^2},\qquad
		\tilde\Gamma_{021}\simeq\Gamma_{021}+4\phi_1+\phi_2.
		\label{eq:rates_dephasing}
	\end{equation}
	Vibrational dephasing enters both rates equally, with the weight $(J/\chi)^2$ of the admixtures, whereas emitter dephasing enters only the phase-loss rate, at full strength. The two rates therefore coincide without emitter dephasing, and emitter dephasing separates them by $4\phi_1+\phi_2$.
	
	This separation determines the three-photon statistics. The conversion of photon triplets into the composite remains coherent for a time $1/\tilde\Gamma_{021}$, and the converted population survives for $1/(2\Gamma_{021})$. The dephased fraction of the converted population returns to the cavity as photon triplets without a fixed phase, and these triplets no longer interfere with the direct three-photon emission. For a composite much narrower than the three-photon state, the three-photon correlation reads [Eq.~\eqref{eq:C_g3}]
	\begin{equation}
		g_-^{(3)}(0)\simeq\frac{1+\eta_3-\tilde\eta_3}{(1+\eta_3)(1+\tilde\eta_3)},
		\label{eq:g3_dephasing}
	\end{equation}
	where $\eta_3=4\zeta_3^2/(\kappa_a\Gamma_{021})$ is the cooperativity of Eq.~\eqref{eq:control_law} and $\tilde\eta_3=4\zeta_3^2/(\kappa_a\tilde\Gamma_{021})$ is that of the coherent conversion. Without emitter dephasing, $\tilde\eta_3=\eta_3$, and Eq.~\eqref{eq:g3_dephasing} reduces to Eq.~\eqref{eq:control_law}. Emitter dephasing lowers the coherent cooperativity and produces the incoherent return of photon triplets. This return grows with the population converted into the composite, and a deeper dip is therefore more sensitive to emitter dephasing. The three-photon correlation doubles its noiseless value once $4\phi_1+\phi_2$ reaches $\Gamma_{021}/\eta_3$. At this rate, the incoherent return equals the coherent residual $(1+\eta_3)^{-2}$ of Eq.~\eqref{eq:control_law} [Eq.~\eqref{eq:C_g3deep}]. To order $(J/\chi)^2$, vibrational dephasing changes only the decay rates and keeps the dip in the form of Eq.~\eqref{eq:control_law}. The next order adds a weak dephasing of the resonant pair [Eq.~\eqref{eq:C_fourth}]. Vibrational dephasing begins to fill the dip once its leakage in Eq.~\eqref{eq:rates_dephasing} reaches the noiseless decay rate of the composite, at $\Gamma_v\simeq2\chi^2\Gamma_{021}/(2J_1^2+J_2^2/4)$. The two noises therefore fill the dip at rates separated by a factor of order $\eta_3(\chi/J)^2$, given to leading order in $J/\chi$ by $2\eta_3\chi^2/(2J_1^2+J_2^2/4)$ [Eqs.~\eqref{eq:C_g3deep} and \eqref{eq:rates_dephasing}].
	
	\begin{figure}
		\centering
		\includegraphics[height=0.43\linewidth]{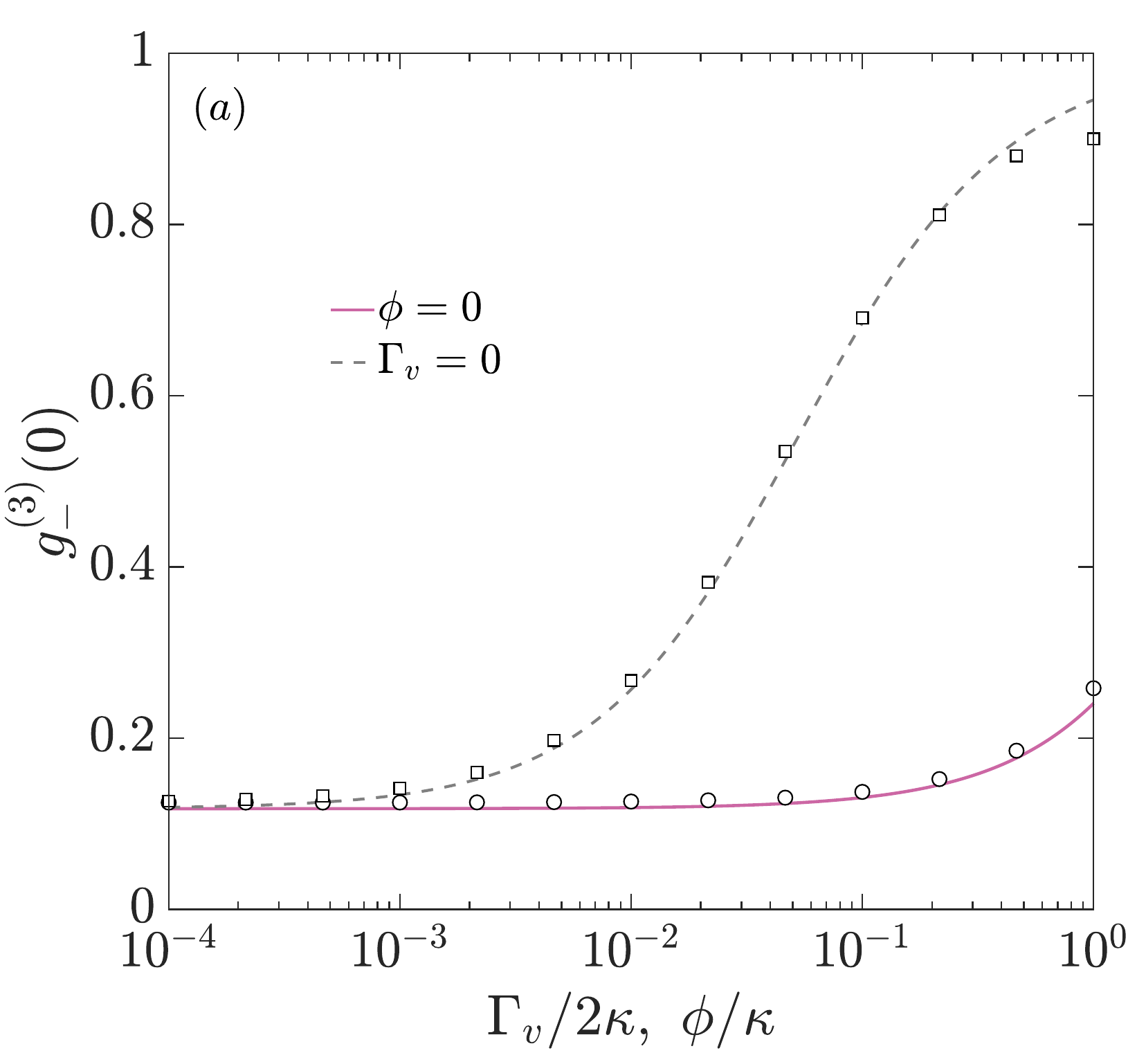}
		\hfill
		\includegraphics[height=0.43\linewidth]{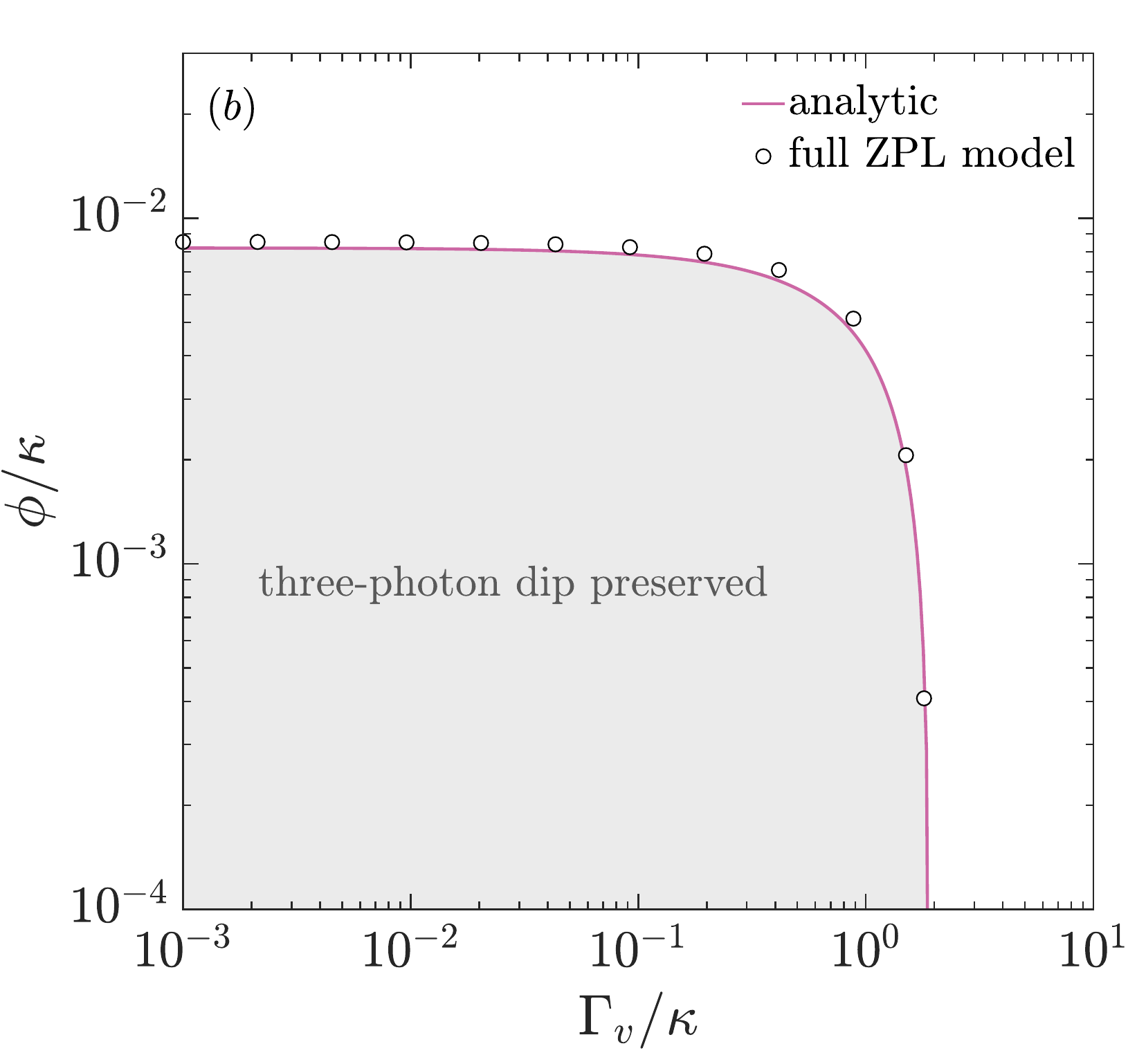}
		\caption{
			(a) Three-photon correlation $g_-^{(3)}(0)$ at the noiseless resonance of Fig.~\ref{fig:threephoton}(b) versus the dephasing rate of a single excitation of subensemble 1, $\Gamma_v/2$ for vibrational dephasing ($\phi_1=\phi_2=0$; solid line and circles) and $\phi_1=\phi_2$ for emitter dephasing ($\Gamma_v=0$; dashed line and squares). (b) Emitter dephasing rate $\phi_1=\phi_2$ at which $g_-^{(3)}(0)$ doubles its noiseless value, versus $\Gamma_v$; in the shaded region, $g_-^{(3)}(0)$ stays below this value. In both panels, lines follow from Eq.~\eqref{eq:C_g3exact} and symbols from the master equation~\eqref{eq:ME}. Parameters as in Fig.~\ref{fig:threephoton}.
		}\label{fig:dephasing}
	\end{figure}
	
	Figure~\ref{fig:dephasing} shows the $s=-1$ resonance of Fig.~\ref{fig:threephoton} under dephasing, at the detunings of the noiseless resonance. Figure~\ref{fig:dephasing}(a) compares the two noises at equal dephasing rates of a single excitation of subensemble 1. The coherence of this excitation with the photonic states is damped at the rate $\Gamma_v/2$ by vibrational dephasing and at the rate $\phi$ by emitter dephasing with $\phi_1=\phi_2=\phi$ [Eq.~\eqref{eq:C_damping}]. Emitter dephasing fills the dip at small rates, whereas vibrational dephasing of the same strength leaves it nearly unchanged up to rates larger by a factor of order $\eta_3(\chi/J)^2$. The master-equation results closely follow Eq.~\eqref{eq:C_g3exact}. This confirms that the shared vibration acts on the resonant pair through the virtual admixtures of Eq.~\eqref{eq:rates_dephasing}.
	
	Figure~\ref{fig:dephasing}(b) shows the combined effect of the two noises through the boundary at which $g_-^{(3)}(0)$ doubles its noiseless value. Along this boundary, the emitter-dephasing threshold $4\phi_1+\phi_2\simeq\Gamma_{021}/\eta_3$ remains nearly unchanged until the vibrational leakage becomes comparable to the decay of the composite. The boundary ends where vibrational dephasing alone doubles the three-photon correlation. Emitter dephasing therefore sets the tighter of the two noise constraints on the three-photon dip. For the ratio $|g_2|/g_1=2+\epsilon$ of Sec.~\ref{sec:threephoton}, the composite retains the displacement $-\epsilon$, and the shared vibration dephases the resonant pair directly at the rate $\Gamma_v\epsilon^2/2$ [Eq.~\eqref{eq:C_damping}], which enters $\tilde\Gamma_{021}$ in the same way as $4\phi_1+\phi_2$. The displacement compensation that creates the three-photon resonance also protects it against the dephasing of the shared vibration.

	\section{Conclusion}\label{sec.5}
	We have shown that when two emitter subensembles displace a common vibration, the relative direction and magnitude of their displacements select the order of resonant multiphoton conversion. For opposite displacements with an integer magnitude ratio, displacement neutrality makes an emitter composite resonant with a multiphoton state, and the vibration-mediated cross interaction makes the conversion pathways between them add constructively. Independent vibrations can restore the resonance, yet their conversion pathways cancel at leading order; the resonance and the conversion thus share a common origin in the shared displacement structure. The same neutrality protects the coherence between the resonant states against direct dephasing by the shared vibration. Energy shifts and pathway interference, the ingredients of conventional and unconventional photon blockade, both enter this mechanism, and the shared displacement structure ties them together: the displacement energies fix the resonance and the signs of the conversion pathways, and displacement neutrality fixes the number of converted photons. Interference that keeps the state with one excitation in each subensemble empty can also make the two-photon correlations of opposite- and same-side displacements coincide at all delays, whereas their three-photon statistics differ.
	
	Our description of the conversion applies within the zero-phonon-line approximation, for subensembles much larger than the number of converted photons and for exchange couplings weak compared with the displacement energies. Deep suppression of $g^{(3)}(0)$ further requires a large conversion cooperativity against cavity and composite decay, together with weak emitter dephasing. Strain coupling to nanomechanical motion has been demonstrated for embedded quantum dots~\cite{2014NatNano106} and proposed for spins in diamond nanostructures~\cite{2013PRL156402}, and the spatial variation of the strain field lets the emitter positions set the relative signs and magnitudes of such couplings. Because the displacement ratio sets the photon number at which the conversion acts, displacement neutrality offers a design rule for tailoring photon correlations of a chosen order. The combination of signed, excitation-dependent displacements of a common bosonic mode and coherent exchange pathways also motivates exploring analogous multiexcitation selection in spin--mechanical architectures~\cite{2010NatPhys602,2013PRL156402}.
	
	\appendix
	\section{Weak-driving hierarchy without dephasing}
	\label{app:weakdrive}
	
	In this Appendix, we derive the weak-driving amplitudes and photon correlations used in Sec.~\ref{sec:correlations}. It answers three questions in turn: where the configuration first enters the excitation hierarchy (Appendix~\ref{app:manifolds}), when the two-excitation response becomes identical for both configurations and why the coincidence extends to the delayed correlation $g_s^{(2)}(\tau)$ (Appendix~\ref{app:screening}), and why this coincidence does not extend to the three-excitation manifold, together with the pathway that carries the configuration selectivity of $g^{(3)}(0)$ (Appendix~\ref{app:three}). The same weak-driving description underlies the rest of the paper: Appendix~\ref{app:dispersive} evaluates it in the dispersive regime, the full weak-driving results of Fig.~\ref{fig:threephoton} are obtained from it directly, and Appendix~\ref{app:dephasing} adds the incoherent populations produced by dephasing.
	
	We consider the ZPL Hamiltonian of Eq.~\eqref{eq:HZPL} with cavity and emitter decay, set $\Gamma_v=\phi_1=\phi_2=0$, and take $\Delta_1=\Delta_2=\Delta_x$. The leading-order photon correlations then follow from the weak-driving recursion of Eq.~\eqref{eq:A_recursion}.
	
	\subsection{Excitation-manifold structure}
	\label{app:manifolds}
	
	Here we organize the weak-driving steady state by excitation number and show that the configuration first enters the photon correlations through the mixed state $|011\rangle$, with one excitation in each subensemble. A Fock state $|n_an_1n_2\rangle$ has the complex energy
	\begin{equation}
		E_{n_an_1n_2}
		=n_a\alpha
		+\sum_{\mu}\Big[n_\mu\beta_\mu
		+\tfrac{1}{2}U_\mu n_\mu(n_\mu-1)\Big]
		+U_{12}n_1n_2,
		\label{eq:A_energy}
	\end{equation}
	where $\alpha=\Delta_a-i\kappa_a$ and $\beta_\mu=\Delta_x-g_\mu^2/\omega_v-i\kappa_x$ are the complex detunings of the cavity and of the polaron-shifted emitter modes, $U_\mu=-2g_\mu^2/\omega_v$ is the interaction energy of two excitations in subensemble $\mu$, and $U_{12}=-2sg_1|g_2|/\omega_v$ is the cross interaction, the only energy that depends on $s$.
	
	The drive adds one excitation at a time, and we write the steady state as $|\psi_s\rangle=\sum_{m\ge0}|\psi_s^{(m)}\rangle$, where $|\psi_s^{(m)}\rangle=\sum_{n_a+n_1+n_2=m}C^{(s)}_{n_an_1n_2}|n_an_1n_2\rangle$ contains the $m$-excitation amplitudes and $C^{(s)}_{000}\simeq1$. The $m$-excitation amplitudes are of order $\Omega^m$ and are driven from the $(m-1)$-excitation manifold,
	\begin{equation}
		H_{{\rm nh},s}^{(m)}|\psi_s^{(m)}\rangle
		=-\Omega\,\hat a^\dagger|\psi_s^{(m-1)}\rangle,
		\label{eq:A_recursion}
	\end{equation}
	where $H_{{\rm nh},s}^{(m)}$ is the undriven non-Hermitian Hamiltonian within the $m$-excitation manifold. The normally ordered photon moments are then $\langle\hat a^{\dagger n}\hat a^n\rangle_s\simeq n!|C_{n00}^{(s)}|^2$, so that
	\begin{equation}
		g_s^{(n)}(0)
		\simeq\frac{n!\,|C_{n00}^{(s)}|^2}{|C_{100}|^{2n}}.
		\label{eq:A_gn}
	\end{equation}
	
	In the one-excitation manifold, the cavity photon $|100\rangle$ exchanges with a single excitation of subensemble 1, $|010\rangle$, or of subensemble 2, $|001\rangle$. In this order,
	\begin{equation}
		H_{\rm nh}^{(1)}=
		\begin{pmatrix}
			\alpha & J_1 & J_2\\
			J_1 & \beta_1 & 0\\
			J_2 & 0 & \beta_2
		\end{pmatrix}.
		\label{eq:A_H1}
	\end{equation}
	A single excitation occupies only one subensemble, so that this manifold contains no $U_{12}$ and is the same for both configurations. The two-excitation manifold contains the two-photon state $|200\rangle$, the states $|110\rangle$ and $|101\rangle$ with one photon and one excitation of subensemble 1 or 2, and the photonless states $|020\rangle$, $|011\rangle$, and $|002\rangle$ with two emitter excitations. In this order,
	\begin{equation}
		H_{{\rm nh},s}^{(2)}=
		\begin{pmatrix}
			E_{200} & \sqrt{2}J_1 & \sqrt{2}J_2 & 0 & 0 & 0\\
			\sqrt{2}J_1 & E_{110} & 0 & \sqrt{2}J_1 & J_2 & 0\\
			\sqrt{2}J_2 & 0 & E_{101} & 0 & J_1 & \sqrt{2}J_2\\
			0 & \sqrt{2}J_1 & 0 & E_{020} & 0 & 0\\
			0 & J_2 & J_1 & 0 & E_{011}^{(s)} & 0\\
			0 & 0 & \sqrt{2}J_2 & 0 & 0 & E_{002}
		\end{pmatrix},
		\label{eq:A_H2}
	\end{equation}
	with the energies of Eq.~\eqref{eq:A_energy}. The mixed state $|011\rangle$ is the only two-excitation state that contains $U_{12}$, through its energy $E_{011}^{(s)}=\beta_1+\beta_2+U_{12}$. The two configurations therefore differ only in the energy of this state,
	\begin{equation}
		H_{{\rm nh},+}^{(2)}-H_{{\rm nh},-}^{(2)}
		=-\frac{4g_1|g_2|}{\omega_v}
		|011\rangle\langle011|.
		\label{eq:A_H2diff}
	\end{equation}
	This shows that the configuration first enters the photon correlations through the two-excitation manifold, and there only through $|011\rangle$.
	
	\subsection{Two-excitation amplitudes and screening}
	\label{app:screening}
	
	Here we determine when the two-excitation response becomes the same for both configurations, and show that the coincidence then extends to the delayed correlation $g_s^{(2)}(\tau)$. For $m=1$, Eq.~\eqref{eq:A_recursion} gives
	\begin{equation}
		C_{100}=-\frac{\Omega}{\Lambda_1},\qquad
		C_{010}=-\frac{J_1}{\beta_1}C_{100},\qquad
		C_{001}=-\frac{J_2}{\beta_2}C_{100},
		\label{eq:A_C1}
	\end{equation}
	where $\Lambda_1=\alpha-J_1^2/\beta_1-J_2^2/\beta_2$ is the complex cavity detuning dressed by exchange with the two emitter modes. This one-excitation response is the same for both configurations and sets the reference for the higher-order correlations.
	
	In the two-excitation manifold, the drive adds a photon to the one-excitation states and thereby feeds $|200\rangle$, $|110\rangle$, and $|101\rangle$. The photonless states $|020\rangle$, $|011\rangle$, and $|002\rangle$ are reached only by exchange. The energies of these states follow from Eq.~\eqref{eq:A_energy}: $E_{200}=2\alpha$, $E_{110}=\alpha+\beta_1$, $E_{101}=\alpha+\beta_2$, $E_{020}=2\beta_1+U_1$, $E_{002}=2\beta_2+U_2$, and $E_{011}^{(s)}=\beta_1+\beta_2+U_{12}$. The mixed state is reached from $|110\rangle$ and $|101\rangle$ along the two exchange pathways of Eq.~\eqref{eq:C011}.
	
	The configuration drops out of the two-excitation response when the two pathways cancel, $J_2C_{110}+J_1C_{101}=0$. The mixed state then remains empty, and the remaining amplitude equations contain no $U_{12}$. In these equations, $|020\rangle$ and $|002\rangle$, with two excitations in one subensemble, shift $|110\rangle$ and $|101\rangle$ through virtual exchange to the dressed energies
	\begin{equation}
		\mathcal E_{110}=E_{110}-\frac{2J_1^2}{E_{020}},
		\qquad
		\mathcal E_{101}=E_{101}-\frac{2J_2^2}{E_{002}}.
		\label{eq:A_dressed}
	\end{equation}
	The cancellation requires $C_{110}/J_1=-C_{101}/J_2$, so that the two pathway amplitudes $J_2C_{110}$ and $J_1C_{101}$ have equal magnitudes and opposite phases. The amplitude equations of $|110\rangle$ and $|101\rangle$ then give
	\begin{equation}
		\frac{C_{110}}{J_1}=-\frac{C_{101}}{J_2}=\frac{\Omega C_{100}}{\mathcal E_{110}+\mathcal E_{101}}
		\left(\frac{1}{\beta_1}-\frac{1}{\beta_2}\right),\qquad
		\frac{\sqrt{2}C_{200}}{\Omega C_{100}}
		=\frac{\beta_1\mathcal E_{110}+\beta_2\mathcal E_{101}}
		{\beta_1\beta_2(\mathcal E_{110}+\mathcal E_{101})}.
		\label{eq:A_ratio}
	\end{equation}
	The pathway amplitudes are proportional to $1/\beta_1-1/\beta_2$, the difference between the responses of the two emitter modes. For equal bare emitter detunings, $\beta_1-\beta_2=-(g_1^2-g_2^2)/\omega_v$. If the two subensembles had equal displacement magnitudes, both pathways would vanish, and the $|200\rangle$ amplitude equation would then require $\alpha=-\beta_1$, which is impossible for positive decay rates. Screening therefore requires $g_1^2\neq g_2^2$. Inserting Eq.~\eqref{eq:A_ratio} into the $|200\rangle$ amplitude equation gives the screening condition
	\begin{equation}
		L=J_1^2A_1+J_2^2A_2,
		\label{eq:A_screen}
	\end{equation}
	with
	\begin{equation}
		L=(\beta_1+\beta_2)E_{110}E_{101},\qquad
		A_1=\beta_1-\beta_2+\frac{2\beta_1E_{101}}{E_{020}},\qquad
		A_2=\beta_2-\beta_1+\frac{2\beta_2E_{110}}{E_{002}}.
		\label{eq:A_LA}
	\end{equation}
	Here $L$ contains the energy $\beta_1+\beta_2$ of the mixed state without its cross interaction and the energies $E_{110}$ and $E_{101}$ of the two states that feed it, and the coefficients $A_1$ and $A_2$ describe the response of subensembles 1 and 2 to the exchange, including virtual exchange with $|020\rangle$ and $|002\rangle$, respectively. Each exchange coupling enters twice, once when an excitation leaves the cavity mode and once when it returns to $|200\rangle$ or proceeds to $|020\rangle$ or $|002\rangle$, so that Eq.~\eqref{eq:A_screen} is linear in $J_1^2$ and $J_2^2$, and none of $L$, $A_1$, and $A_2$ contains $U_{12}$. With $|011\rangle$ empty, the drive fixes the amplitudes of $|200\rangle$, $|110\rangle$, and $|101\rangle$; Eq.~\eqref{eq:A_screen} is the additional requirement that these amplitudes also cancel the two pathways into $|011\rangle$. As a complex condition, it balances both the magnitudes and the phases of the two pathways and thereby fixes the two exchange couplings at given detunings and decay rates,
	\begin{equation}
		J_1^2=\frac{{\rm Im}(LA_2^*)}{{\rm Im}(A_1A_2^*)},\qquad
		J_2^2=\frac{{\rm Im}(A_1L^*)}{{\rm Im}(A_1A_2^*)}.
		\label{eq:A_Jscreen}
	\end{equation}
	Without losses, $L$, $A_1$, and $A_2$ are real, and Eq.~\eqref{eq:A_screen} reduces to a single condition, which balances the magnitudes of the two pathways and leaves a one-parameter family of couplings. The finite lifetimes of the intermediate states give the two pathways phases of first order in the decay rates, and the imaginary part of Eq.~\eqref{eq:A_screen} requires these phases to coincide. Both ${\rm Im}(A_1A_2^*)$ and the numerators in Eq.~\eqref{eq:A_Jscreen} are therefore proportional to the decay rates, so that for weak losses the exchange couplings depend on the decay rates only through $\kappa_x/\kappa_a$.
	The solutions with positive $J_1^2$ and $J_2^2$ form the screened branch of Sec.~\ref{sec:correlations}. On this branch, the amplitudes of Eq.~\eqref{eq:A_ratio} satisfy Eq.~\eqref{eq:C011} with $C_{011}=0$ for either sign of $U_{12}$. Since every two-excitation state decays, these driven amplitudes are fixed uniquely by their source. This shows that all leading-order two-excitation amplitudes are identical for $s=\pm1$, although $U_{12}$ is finite.
	
	Away from the screened branch, the configuration enters the two-photon correlation through $|011\rangle$. Eliminating first the photonless states $|020\rangle$, $|011\rangle$, and $|002\rangle$ and then $|110\rangle$ and $|101\rangle$ gives Eq.~\eqref{eq:g2_weak}, with
	\begin{align}
		\Sigma_{2,s}&=\frac{J_1^2M_{2,s}+J_2^2M_{1,s}-2J_1J_2K_s}{D_s},
		\nonumber\\
		\mathcal R_{2,s}&=\frac{J_1^2M_{2,s}/\beta_1+J_2^2M_{1,s}/\beta_2
			-J_1J_2K_s(\beta_1^{-1}+\beta_2^{-1})}{D_s}.
		\label{eq:A_SigmaR}
	\end{align}
	Here $M_{1,s}=\mathcal E_{110}-J_2^2/E_{011}^{(s)}$ and $M_{2,s}=\mathcal E_{101}-J_1^2/E_{011}^{(s)}$ are the energies of $|110\rangle$ and $|101\rangle$ dressed by exchange with the photonless states, $K_s=-J_1J_2/E_{011}^{(s)}$ is their coupling through $|011\rangle$, and $D_s=M_{1,s}M_{2,s}-K_s^2$ becomes small when a dressed state formed by $|110\rangle$ and $|101\rangle$ is resonant with the drive. These quantities, and hence $\Sigma_{2,s}$ and $\mathcal R_{2,s}$, depend on $s$ only through $E_{011}^{(s)}$. On the screened branch, Eqs.~\eqref{eq:A_C1} and \eqref{eq:A_ratio} give
	\begin{equation}
		g^{(2)}(0)
		=|\Lambda_1|^2
		\frac{|\beta_1\mathcal E_{110}+\beta_2\mathcal E_{101}|^2}
		{|\beta_1\beta_2(\mathcal E_{110}+\mathcal E_{101})|^2},
		\label{eq:A_g2_screen}
	\end{equation}
	which contains no $U_{12}$.
	
	The delayed correlation $g_s^{(2)}(\tau)$ of Sec.~\ref{sec:correlations} is set by the state of the system after a photon detection and by its subsequent evolution~\cite{1991OC73}. A detection at $\tau=0$ leaves the system in
	\begin{equation}
		\hat a|\psi_s\rangle
		=C_{100}|000\rangle+\sqrt{2}C_{200}^{(s)}|100\rangle
		+C_{110}^{(s)}|010\rangle+C_{101}^{(s)}|001\rangle,
		\label{eq:A_cond}
	\end{equation}
	to leading order in $\Omega$. The system then evolves within the $s$-independent one-excitation manifold of Eq.~\eqref{eq:A_H1}, which the drive refills from the vacuum. Since the mixed state $|011\rangle$ contains no photon, the configuration enters the conditional state only through $C_{200}^{(s)}$, $C_{110}^{(s)}$, and $C_{101}^{(s)}$, which are identical for $s=\pm1$ on the screened branch. This shows that $g_+^{(2)}(\tau)=g_-^{(2)}(\tau)$ at all delays.
	
	\subsection{Three-excitation response}
	\label{app:three}
	
	This subsection shows why the screening does not extend to the three-excitation manifold and identifies the dominant configuration-sensitive pathway in $g^{(3)}(0)$: at the point of Fig.~\ref{fig:screening}(a), the selectivity is set mainly by the effective energy of $|111\rangle$, which contains one photon and one excitation in each subensemble, relative to an interference zero of the three-photon amplitude. The three-photon correlation requires the next step of the hierarchy. For $m=3$, Eq.~\eqref{eq:A_recursion} involves the ten states with $n_a+n_1+n_2=3$. As in the two-excitation manifold, the drive adds a cavity photon to the two-excitation states and thereby feeds only states that contain at least one photon, whereas $|030\rangle$, $|021\rangle$, $|012\rangle$, and $|003\rangle$ have no direct drive source. The configuration now enters through the three mixed states, in which both subensembles are excited. From Eq.~\eqref{eq:A_energy}, their energies are
	\begin{equation}
		E_{111}^{(s)}=\alpha+\beta_1+\beta_2+U_{12},\qquad
		E_{021}^{(s)}=2\beta_1+\beta_2+U_1+2U_{12},\qquad
		E_{012}^{(s)}=\beta_1+2\beta_2+U_2+2U_{12},
		\label{eq:A_E3}
	\end{equation}
	so that the cross interaction appears once in $|111\rangle$ and twice in the photonless states $|021\rangle$ and $|012\rangle$. These states are populated through
	\begin{align}
		E_{111}^{(s)}C_{111}^{(s)}+\sqrt2\left(J_2C_{210}^{(s)}+J_1C_{201}^{(s)}\right)+\sqrt2\left(J_1C_{021}^{(s)}+J_2C_{012}^{(s)}\right)&=-\Omega C_{011}^{(s)},
		\nonumber\\
		E_{021}^{(s)}C_{021}^{(s)}+J_2C_{120}^{(s)}+\sqrt2J_1C_{111}^{(s)}&=0,
		\nonumber\\
		E_{012}^{(s)}C_{012}^{(s)}+J_1C_{102}^{(s)}+\sqrt2J_2C_{111}^{(s)}&=0.
		\label{eq:A_feed3}
	\end{align}
	Like $|011\rangle$, which is fed from $|110\rangle$ and $|101\rangle$, the mixed state $|111\rangle$ is fed by exchange from $|210\rangle$ and $|201\rangle$, which contain two photons and one emitter excitation; it is additionally driven from $|011\rangle$. The photonless states $|021\rangle$ and $|012\rangle$ are reached from $|111\rangle$ and from $|120\rangle$ and $|102\rangle$, which contain one photon and two excitations of the same subensemble and are fed by the drive from $|020\rangle$ and $|002\rangle$.
	
	On the screened branch, the two-excitation amplitudes are identical for $s=\pm1$ and $C_{011}=0$. The source $-\Omega\hat a^\dagger|\psi_s^{(2)}\rangle$ of the three-excitation manifold is therefore the same for both configurations, and the drive of $|111\rangle$ from $|011\rangle$ in Eq.~\eqref{eq:A_feed3} vanishes. The three-excitation amplitudes of the two configurations then obey the same equations with the same source, except for the energies of the mixed states,
	\begin{equation}
		H_{{\rm nh},+}^{(3)}-H_{{\rm nh},-}^{(3)}=-\frac{4g_1|g_2|}{\omega_v}\Big(|111\rangle\langle111|+2|021\rangle\langle021|+2|012\rangle\langle012|\Big).
		\label{eq:A_H3diff}
	\end{equation}
	Exact screening of the three-excitation response would require these three states to remain empty. By Eq.~\eqref{eq:A_feed3}, this would in turn require $C_{120}=C_{102}=0$ and $J_2C_{210}+J_1C_{201}=0$. The states $|120\rangle$ and $|102\rangle$ are driven directly from $|020\rangle$ and $|002\rangle$, which are populated on the screened branch. They remain empty only if the exchange from $|210\rangle$ and $|201\rangle$ cancels this drive, $2J_1C_{210}=-\Omega C_{020}$ and $2J_2C_{201}=-\Omega C_{002}$. With the amplitudes $C_{020}=-\sqrt2J_1C_{110}/E_{020}$ and $C_{002}=-\sqrt2J_2C_{101}/E_{002}$ of Eq.~\eqref{eq:A_H2} and the screening relation $J_2C_{110}+J_1C_{101}=0$, the remaining condition $J_2C_{210}+J_1C_{201}=0$ holds only if $|020\rangle$ and $|002\rangle$ are degenerate, $E_{020}=E_{002}$. For equal bare emitter detunings, $|E_{020}-E_{002}|$ equals the difference of displacement energies in Eq.~\eqref{eq:mismatch}, which is finite for $g_1^2\neq g_2^2$. The configuration therefore enters the three-photon response through the energies of $|111\rangle$, $|021\rangle$, and $|012\rangle$.
	
	To obtain an explicit expression for the three-photon response, we omit the configuration label $s$ from here on; every quantity that contains $E_{111}^{(s)}$, $E_{021}^{(s)}$, or $E_{012}^{(s)}$ depends on the configuration, and $C_{011}=0$ on the screened branch. We first eliminate the four photonless states $|030\rangle$, $|021\rangle$, $|012\rangle$, and $|003\rangle$. Their virtual exchange with $|120\rangle$, $|111\rangle$, and $|102\rangle$ gives the effective amplitude equations
	\begin{equation}
		\begin{pmatrix}
			\vartheta_{120} & \vartheta_{120,111} & 0\\
			\vartheta_{120,111} & \vartheta_{111} & \vartheta_{111,102}\\
			0 & \vartheta_{111,102} & \vartheta_{102}
		\end{pmatrix}
		\begin{pmatrix}
			C_{120}\\
			C_{111}\\
			C_{102}
		\end{pmatrix}
		=
		-\begin{pmatrix}
			2J_1C_{210}+\Omega C_{020}\\
			\sqrt{2}\left(J_2C_{210}+J_1C_{201}\right)\\
			2J_2C_{201}+\Omega C_{002}
		\end{pmatrix},
		\label{eq:A_effective3}
	\end{equation}
	where
	\begin{align}
		\vartheta_{120}&=E_{120}-\frac{3J_1^2}{E_{030}}-\frac{J_2^2}{E_{021}},\qquad
		\vartheta_{102}=E_{102}-\frac{J_1^2}{E_{012}}-\frac{3J_2^2}{E_{003}},
		\nonumber\\
		\vartheta_{111}&=E_{111}-\frac{2J_1^2}{E_{021}}-\frac{2J_2^2}{E_{012}},
		\nonumber\\
		\vartheta_{120,111}&=-\frac{\sqrt{2}J_1J_2}{E_{021}},\qquad
		\vartheta_{111,102}=-\frac{\sqrt{2}J_1J_2}{E_{012}}.
		\label{eq:A_vartheta3}
	\end{align}
	Here $E_{120}=\alpha+2\beta_1+U_1$, $E_{102}=\alpha+2\beta_2+U_2$, $E_{030}=3\beta_1+3U_1$, and $E_{003}=3\beta_2+3U_2$, while $E_{111}$, $E_{021}$, and $E_{012}$ are given in Eq.~\eqref{eq:A_E3}. The diagonal elements $\vartheta_{120}$, $\vartheta_{111}$, and $\vartheta_{102}$ are the energies of $|120\rangle$, $|111\rangle$, and $|102\rangle$ dressed by virtual exchange with the photonless states. The off-diagonal element $\vartheta_{120,111}$ is the effective exchange coupling between $|120\rangle$ and $|111\rangle$ through virtual population of $|021\rangle$, and $\vartheta_{111,102}$ couples $|111\rangle$ and $|102\rangle$ through $|012\rangle$.
	
	In the remaining six states, the three-photon state $|300\rangle$ exchanges with $|210\rangle$ and $|201\rangle$. These exchange, in turn, with $|120\rangle$ and $|102\rangle$, respectively, and both exchange with the mixed state $|111\rangle$, which thus links the subensemble-1 pair $|210\rangle$, $|120\rangle$ to the subensemble-2 pair $|201\rangle$, $|102\rangle$. Eliminating the two pairs leaves two coupled amplitudes, those of $|300\rangle$ and $|111\rangle$, which we refer to as the two channels.
	
	The amplitude equations of $|210\rangle$ and $|201\rangle$ are
	\begin{align}
		E_{210}C_{210}+\sqrt3J_1C_{300}+2J_1C_{120}+\sqrt2J_2C_{111}&=-\sqrt2\,\Omega C_{110},
		\nonumber\\
		E_{201}C_{201}+\sqrt3J_2C_{300}+2J_2C_{102}+\sqrt2J_1C_{111}&=-\sqrt2\,\Omega C_{101},
		\label{eq:A_feed210}
	\end{align}
	with $E_{210}=2\alpha+\beta_1$ and $E_{201}=2\alpha+\beta_2$. Eliminating $|120\rangle$ and $|102\rangle$ with the first and third lines of Eq.~\eqref{eq:A_effective3} gives
	\begin{align}
		\mathcal E_{210}C_{210}+\sqrt3J_1C_{300}+\sqrt2J_2\xi_{210}C_{111}&=-\Omega\mathcal S_{210},
		\nonumber\\
		\mathcal E_{201}C_{201}+\sqrt3J_2C_{300}+\sqrt2J_1\xi_{201}C_{111}&=-\Omega\mathcal S_{201},
		\label{eq:A_pairs}
	\end{align}
	where
	\begin{align}
		\mathcal E_{210}&=E_{210}-\frac{4J_1^2}{\vartheta_{120}},\qquad
		\xi_{210}=1+\frac{2J_1^2}{\vartheta_{120}E_{021}},\qquad
		\mathcal S_{210}=\sqrt2\,C_{110}-\frac{2J_1C_{020}}{\vartheta_{120}},
		\nonumber\\
		\mathcal E_{201}&=E_{201}-\frac{4J_2^2}{\vartheta_{102}},\qquad
		\xi_{201}=1+\frac{2J_2^2}{\vartheta_{102}E_{012}},\qquad
		\mathcal S_{201}=\sqrt2\,C_{101}-\frac{2J_2C_{002}}{\vartheta_{102}}.
		\label{eq:A_pair_coeff}
	\end{align}
	Here $\mathcal E_{210}$ is the energy of $|210\rangle$ dressed by exchange with $|120\rangle$, which is itself dressed by the photonless states. The term $-\Omega\mathcal S_{210}$ is the total drive of $|210\rangle$: the drive from $|110\rangle$ and the drive that $|120\rangle$ receives from $|020\rangle$ and passes to $|210\rangle$ by exchange. The factor $\xi_{210}$ is the amplitude for $|210\rangle$ to reach $|111\rangle$ in units of the direct exchange in subensemble 2; its second term is the pathway through $|120\rangle$ and $|021\rangle$. The quantities $\mathcal E_{201}$, $\mathcal S_{201}$, and $\xi_{201}$ describe $|201\rangle$ in the same way, with the roles of the two subensembles interchanged.
	
	Using Eqs.~\eqref{eq:A_effective3} and \eqref{eq:A_pairs} to eliminate $|120\rangle$, $|102\rangle$, $|210\rangle$, and $|201\rangle$ from the amplitude equation of $|300\rangle$, $E_{300}C_{300}+\sqrt3J_1C_{210}+\sqrt3J_2C_{201}=-\sqrt3\,\Omega C_{200}$ with $E_{300}=3\alpha$, and from the second line of Eq.~\eqref{eq:A_effective3} gives
	\begin{equation}
		\begin{pmatrix}
			\mathcal E_{300} & \mathcal V\\
			\mathcal V & \mathcal E_{111}
		\end{pmatrix}
		\begin{pmatrix}
			C_{300}\\
			C_{111}
		\end{pmatrix}
		=-\Omega
		\begin{pmatrix}
			\mathcal S_{300}\\
			\mathcal S_{111}
		\end{pmatrix}.
		\label{eq:A_twochannel}
	\end{equation}
	The dressed energies and the effective coupling of the two channels are
	\begin{align}
		\mathcal E_{300}&=E_{300}-\frac{3J_1^2}{\mathcal E_{210}}-\frac{3J_2^2}{\mathcal E_{201}},
		\nonumber\\
		\mathcal E_{111}&=\vartheta_{111}-\frac{\vartheta_{120,111}^2}{\vartheta_{120}}-\frac{\vartheta_{111,102}^2}{\vartheta_{102}}-\frac{2J_2^2\xi_{210}^2}{\mathcal E_{210}}-\frac{2J_1^2\xi_{201}^2}{\mathcal E_{201}},
		\nonumber\\
		\mathcal V&=-\sqrt6\,J_1J_2\left(\frac{\xi_{210}}{\mathcal E_{210}}+\frac{\xi_{201}}{\mathcal E_{201}}\right),
		\label{eq:A_twochannel_E}
	\end{align}
	and the drives of the two channels are
	\begin{align}
		\mathcal S_{300}&=\sqrt3\left(C_{200}-\frac{J_1\mathcal S_{210}}{\mathcal E_{210}}-\frac{J_2\mathcal S_{201}}{\mathcal E_{201}}\right),
		\nonumber\\
		\mathcal S_{111}&=-\frac{\vartheta_{120,111}C_{020}}{\vartheta_{120}}-\frac{\vartheta_{111,102}C_{002}}{\vartheta_{102}}-\sqrt2\left(\frac{J_2\xi_{210}\mathcal S_{210}}{\mathcal E_{210}}+\frac{J_1\xi_{201}\mathcal S_{201}}{\mathcal E_{201}}\right).
		\label{eq:A_twochannel_S}
	\end{align}
	Here $\mathcal E_{300}$ is the energy of the three-photon state dressed by exchange through $|210\rangle$ and $|201\rangle$, and $\mathcal E_{111}$ is the effective energy of $|111\rangle$, dressed by the photonless states and by both pairs. The coupling $\mathcal V$ is the effective exchange between $|300\rangle$ and $|111\rangle$, the sum of the pathways through $|210\rangle$ and through $|201\rangle$. The terms $-\Omega\mathcal S_{300}$ and $-\Omega\mathcal S_{111}$ are the total drives of the two channels. The three-photon state is driven directly from $|200\rangle$ and indirectly through $|210\rangle$ and $|201\rangle$, whereas $|111\rangle$ receives only the drives passed on from $|120\rangle$, $|102\rangle$, $|210\rangle$, and $|201\rangle$.
	
	Solving Eq.~\eqref{eq:A_twochannel} for $C_{300}$ and using Eqs.~\eqref{eq:A_gn} and \eqref{eq:A_C1}, we obtain Eq.~\eqref{eq:g3_twochannel}, with
	\begin{equation}
		\mathcal Z=\frac{\mathcal V\mathcal S_{111}}{\mathcal S_{300}},\qquad
		\mathcal P=\frac{\mathcal V^2}{\mathcal E_{300}},\qquad
		g_{{\rm bg},s}^{(3)}=\frac{6|\Lambda_1|^2|\mathcal S_{300}|^2}{|\mathcal E_{300}|^2|C_{100}|^4}.
		\label{eq:A_g3}
	\end{equation}
	The three-photon amplitude is the sum of a direct pathway and an indirect pathway through $|111\rangle$. The background $g_{{\rm bg},s}^{(3)}$ is the three-photon correlation without the indirect pathway, $\mathcal E_{111}\to\infty$, and is independent of the drive, since $\mathcal S_{300}$ is built from two-excitation amplitudes, which scale as $C_{100}^2$. The interference factor $|(\mathcal E_{111}-\mathcal Z)/(\mathcal E_{111}-\mathcal P)|^2$ measures how close $|111\rangle$ is to the cancellation of the two pathways, which occurs at the interference zero $\mathcal Z$ set by the drives $\mathcal S_{300}$ and $\mathcal S_{111}$ of the two channels and by their coupling $\mathcal V$.
	
	The two configurations differ in Eq.~\eqref{eq:g3_twochannel} through $E_{111}^{(s)}$, which enters $\mathcal E_{111}$ directly, and through $E_{021}^{(s)}$ and $E_{012}^{(s)}$, which enter the dressing of all states. At the point of Fig.~\ref{fig:screening}(a), the background is nearly configuration independent, $g_{{\rm bg},-}^{(3)}=5.12$ and $g_{{\rm bg},+}^{(3)}=4.83$, so that the configuration acts mainly through $\mathcal E_{111}$. For $s=-1$, $\mathcal E_{111}=(6.16-0.08i)\chi$ lies close to $\mathcal Z=(6.76+1.34i)\chi$, the interference factor equals $0.061$, and $g_-^{(3)}(0)=0.31$. For $s=+1$, $\mathcal E_{111}=(-3.66-0.06i)\chi$ lies far from $\mathcal Z=(4.99+1.20i)\chi$, the interference factor equals $0.46$, and $g_+^{(3)}(0)=2.23$. Assigning the $s=-1$ energy $E_{111}^{(-)}$ to $|111\rangle$ alone within the $s=+1$ configuration moves $\mathcal E_{111}$ to $(4.34-0.06i)\chi$ and gives $g^{(3)}(0)=0.32$. This shows that, at the point of Fig.~\ref{fig:screening}(a), the configuration-sensitive indirect pathway through $|111\rangle$ dominates the three-photon interference, while $|021\rangle$ and $|012\rangle$ contribute through the dressing of the two channels.
	
	\section{Dispersive description of the three-photon resonance}
	\label{app:dispersive}
	
	In this Appendix, we derive the dispersive description of the three-photon resonance of Sec.~\ref{sec:threephoton}. It answers four questions in turn: where the resonance lies once virtual cavity--emitter exchange, in which the system passes briefly through off-resonant states and returns, is included (Appendix~\ref{app:lamb}), why it reshapes the three-photon response only for the shared vibration with $s=-1$ and how this extends to integer displacement ratios (Appendix~\ref{app:zeta3}), why the two-photon correlation remains close to unity (Appendix~\ref{app:kerr}), and how deep the resulting dip in $g^{(3)}(0)$ is (Appendix~\ref{app:corr}).
	
	We work within the collective bosonic ZPL model of Eq.~\eqref{eq:HZPL} and take $|g_2|=2g_1$. Single excitations of subensembles 1 and 2 then carry the polaron shifts $-\chi$ and $-4\chi$, with $\chi=g_1^2/\omega_v$, and the interaction energies are $U_1=-2\chi$, $U_2=-8\chi$, and $U_{12}=-4s\chi$. We compare three cases: the shared vibration with $s=-1$ ($U_{12}=4\chi$), the shared vibration with $s=+1$ ($U_{12}=-4\chi$), and two independent vibrations with the same polaron shifts but $U_{12}=0$. For the shared vibration, the displacement energies combine into $-\chi Q^2$, where $Q_\nu=n_1+2sn_2$ is the value of $\hat Q$ of Eq.~\eqref{eq:HP} in the Fock state $|\nu\rangle=|n_an_1n_2\rangle$; for independent vibrations, they reduce to $-\chi(n_1^2+4n_2^2)$. The state of interest is the composite $|021\rangle$, with two excitations in subensemble 1 and one in subensemble 2, which can be brought into resonance with the three-photon state $|300\rangle$; we refer to the two as the resonant pair. Energies are measured in the frame rotating at the drive frequency, in which $\Delta_a$ and $\Delta_x$ are the cavity and common emitter detunings ($\Delta_1=\Delta_2=\Delta_x$), so that an $n$-excitation state resonant with $n$ drive quanta has zero energy, and $E_\nu$ denotes the energy of $|\nu\rangle$ given by Eq.~\eqref{eq:A_energy}. Appendices~\ref{app:lamb}--\ref{app:kerr} determine the energies and couplings of the lossless system, Appendix~\ref{app:corr} adds the cavity and emitter losses, with amplitude decay rates $\kappa_a$ and $\kappa_x$, and Appendix~\ref{app:dephasing} adds pure dephasing. Throughout, $J$ denotes the common scale of $J_1$ and $J_2$ at a fixed ratio $J_2/J_1$, and we consider the dispersive regime $J_\mu\ll\chi$.
	
	\subsection{Dispersive elimination and resonance compensation}
	\label{app:lamb}
	
	The three-photon resonance requires the composite $|021\rangle$ and the three-photon state $|300\rangle$ to remain degenerate once the Lamb shifts produced by virtual cavity--emitter exchange are included. Here we derive these shifts and the detunings that compensate them for the three cases. For $s=-1$, the compensating detunings are of order $J^2/\chi$, whereas independent vibrations require the emitter--cavity detuning $\Delta_x-\Delta_a\simeq8\chi/3$; for $s=+1$, the composite remains $16\chi$ below $|300\rangle$.
	
	Within a given excitation manifold, we divide the states into a retained set $P$, which contains the states that can be brought into resonance with the drive, and the remaining off-resonant states $R$, which are detuned from them by energies of order $\chi$ and are therefore populated only virtually. In the three-excitation manifold, the states of $R$ reached from the resonant pair by a single exchange are the intermediate states of Fig.~\ref{fig:paths}(a). Correspondingly, we split the undriven Hamiltonian $H$ of Eq.~\eqref{eq:HZPL} into $H_{PP}$, which contains the energies $E_\nu$ of the retained states and their direct couplings, $H_{RR}$, which acts within $R$, and $H_{PR}=H_{RP}^\dagger$, which contains the cavity--emitter exchange connecting the two sets. Eliminating $R$ at the drive frequency~\cite{1992Wiley,2012PRA032111} gives
	\begin{equation}
		H_{\rm eff}^{P}=H_{PP}-H_{PR}\,H_{RR}^{-1}H_{RP},
		\label{eq:B_Heff}
	\end{equation}
	where the second term describes virtual excursions from the retained states into $R$ and back. Because the off-resonant states lie far from the drive, these excursions can be organized by the number of cavity--emitter exchanges they involve,
	\begin{equation}
		\langle\nu|H_{\rm eff}^{P}|\nu'\rangle=\langle\nu|H_{PP}|\nu'\rangle+\sum_{\rm paths}\frac{\langle\nu|H|q_1\rangle\langle q_1|H|q_2\rangle\cdots\langle q_\ell|H|\nu'\rangle}{(-E_{q_1})(-E_{q_2})\cdots(-E_{q_\ell})}.
		\label{eq:B_pathsum}
	\end{equation}
	In each virtual process, the system leaves the retained state $|\nu'\rangle$ by exchanging an excitation between the cavity and an emitter mode, passes through the off-resonant states $|q_1\rangle,\ldots,|q_\ell\rangle$, whose energies $E_{q_i}$ measured from the drive follow from Eq.~\eqref{eq:A_energy}, and ends in $|\nu\rangle$. The process grows with the exchange strengths it involves and is suppressed by the detuning of every off-resonant state it visits, so that each additional exchange costs a factor of order $J/\chi$. Processes with two exchanges shift the energies of the retained states, three exchanges convert $|300\rangle$ into the composite (Appendix~\ref{app:zeta3}), and four exchanges in which two virtual emitter excitations are present at the same time and interact produce the residual two-photon shift (Appendix~\ref{app:kerr}). As in Appendix~\ref{app:weakdrive}, we denote the effective diagonal energies by $\mathcal E_\nu=\langle\nu|H_{\rm eff}^{P}|\nu\rangle$; they reduce to $E_\nu$ when the virtual exchange is neglected.
	
	In the one-excitation manifold, $P=\{|100\rangle\}$, and the virtual excitation of the two emitter modes, which lie $\chi-\Delta_x$ and $4\chi-\Delta_x$ below the drive, gives the dressed photon energy, which includes the Lamb shifts of the photon,
	\begin{equation}
		\mathcal E_{100}=\Delta_a+\frac{J_1^2}{\chi-\Delta_x}+\frac{J_2^2}{4\chi-\Delta_x}.
		\label{eq:B_E100}
	\end{equation}
	The last two terms are these Lamb shifts. Since a single excitation does not involve $U_{12}$, Eq.~\eqref{eq:B_E100} is the same in all three cases. In the three-excitation manifold, we retain $P=\{|300\rangle,|021\rangle\}$. Since the effective energy of $|300\rangle$ equals $3\mathcal E_{100}$ up to the residual two-photon shift of Eq.~\eqref{eq:B_Kerr}, we require the dressed photon and the dressed composite to be resonant with the drive, $\mathcal E_{100}=\mathcal E_{021}=0$, before their mutual coupling is included.
	
	\emph{Shared vibration.} For $s=-1$, the composite has $Q_{021}=0$ and bare energy $E_{021}=3\Delta_x$. Its virtual exchange with $|111\rangle$ and $|120\rangle$, which lie $\chi$ and $4\chi$ below it and are reached with the exchange strengths $\sqrt2J_1$ and $J_2$, raises it to
	\begin{equation}
		\mathcal E_{021}\simeq3\Delta_x+\frac{2J_1^2}{\chi}+\frac{J_2^2}{4\chi}.
		\label{eq:B_E021}
	\end{equation}
	Since both detunings are of order $J^2/\chi$, Eq.~\eqref{eq:B_E100} is evaluated at $\Delta_x=0$ to leading order, as in Eq.~\eqref{eq:B_E021}. The conditions $\mathcal E_{100}=0$ and $\mathcal E_{021}=0$ then give $\Delta_a=-(J_1^2+J_2^2/4)/\chi$ and $\Delta_x=-(2J_1^2+J_2^2/4)/(3\chi)$, the detunings of Eq.~\eqref{eq:resonance_compensation}. The corrections to these detunings are of order $J^4/\chi^3$ and shift $\mathcal E_{021}$ by an amount smaller than the three-photon coupling $V_3$ of Eq.~\eqref{eq:B_paths} by a factor $J/\chi$, so that, with the leading-order detunings used in Fig.~\ref{fig:threephoton}(a), the composite remains resonant on the scale of the conversion coupling. For $s=+1$, the same detunings leave the dressed photon energy unchanged, whereas the composite has $Q_{021}=4$ and lies at $E_{021}\simeq-16\chi$. It then belongs to $R$ and is eliminated together with the other off-resonant states.
	
	\emph{Independent vibrations.} For $U_{12}=0$, the composite retains the displacement energy of its three excitations, $E_{021}=3\Delta_x-8\chi$, so that resonance requires $\Delta_x\simeq8\chi/3$. The states $|111\rangle$ and $|120\rangle$ then lie $\chi/3$ and $4\chi/3$ above the composite and lower it, $\mathcal E_{021}\simeq3\Delta_x-8\chi-6J_1^2/\chi-3J_2^2/(4\chi)$, while in Eq.~\eqref{eq:B_E100} the subensemble-1 level now lies above the photon, which reverses the sign of its Lamb shift. The two conditions give
	\begin{equation}
		\Delta_a\simeq\frac{3}{\chi}\left(\frac{J_1^2}{5}-\frac{J_2^2}{4}\right),\qquad
		\Delta_x\simeq\frac{8\chi}{3}+\frac{2J_1^2+J_2^2/4}{\chi}.
		\label{eq:B_detunings_ind}
	\end{equation}
	The emitter detuning must thus supply the displacement energy $8\chi$ of the composite, which for the shared vibration with $s=-1$ is cancelled by the cross interaction $2U_{12}$.
	Because $|111\rangle$ lies only $\chi/3$ from the composite, the corrections to Eq.~\eqref{eq:B_detunings_ind} are larger than for the shared vibration. The detunings of the independent-vibration curve in Fig.~\ref{fig:threephoton}(a) are therefore obtained from the exact conditions $\mathcal E_{100}=\mathcal E_{021}=0$ with Eq.~\eqref{eq:B_Heff}.
	
	\subsection{Three-photon conversion and the role of the shared vibration}
	\label{app:zeta3}
	\label{app:shared}
	
	Bringing the composite into resonance with $|300\rangle$ reshapes the three-photon response only if the two states are coupled. Here we derive this coupling, show that it is generated by the cross interaction $U_{12}$ of the shared vibration, and evaluate it for the three cases: the three conversion paths add for $s=-1$ and cancel for independent vibrations. The same path structure holds for every integer displacement ratio, for which the shared vibration converts $r+1$ photons with the closed-form strength of Eq.~\eqref{eq:zeta_r} and the paths of independent vibrations again cancel at resonance.
	
	To compare the conversion in the three cases, we retain $P=\{|300\rangle,|021\rangle\}$ also for $s=+1$, where the composite is off resonance, and define the amplitude for converting three cavity photons into the composite, $V_3=\langle021|H_{\rm eff}^{P}|300\rangle$. To leading order, it arises from the three-exchange processes of Eq.~\eqref{eq:B_pathsum}, which follow the three paths of Fig.~\ref{fig:paths},
	\begin{equation}
		\begin{aligned}
			&\text{path 1:}\quad|300\rangle\xrightarrow{\sqrt3J_1}|210\rangle\xrightarrow{2J_1}|120\rangle\xrightarrow{J_2}|021\rangle,\\
			&\text{path 2:}\quad|300\rangle\xrightarrow{\sqrt3J_1}|210\rangle\xrightarrow{\sqrt2J_2}|111\rangle\xrightarrow{\sqrt2J_1}|021\rangle,\\
			&\text{path 3:}\quad|300\rangle\xrightarrow{\sqrt3J_2}|201\rangle\xrightarrow{\sqrt2J_1}|111\rangle\xrightarrow{\sqrt2J_1}|021\rangle.
		\end{aligned}
		\label{eq:B_threepaths}
	\end{equation}
	Each arrow carries the exchange amplitude $\langle\nu|H|\nu'\rangle$ of the step $|\nu'\rangle\to|\nu\rangle$. Each path removes the same three photons and creates the same three emitter excitations, only in a different order, so that the bosonic enhancement factors, $\sqrt3\sqrt2$ for removing the photons and $\sqrt2$ for placing two excitations in subensemble 1, are the same on every path. All three paths thus carry the strength $2\sqrt3J_1^2J_2$ and differ only in the energies $d_\nu=E_\nu-E_{300}$ of the intermediate states they visit. Since $E_{300}\simeq0$ at resonance, the detuning of each intermediate state from the drive equals its energy $d_\nu$ measured from the resonant pair. Summing the three contributions gives
	\begin{equation}
		V_3\simeq2\sqrt3J_1^2J_2\left(\frac{1}{d_{210}d_{120}}+\frac{1}{d_{210}d_{111}}+\frac{1}{d_{201}d_{111}}\right).
		\label{eq:B_paths}
	\end{equation}
	Each path is suppressed by the energies of the two intermediate states it visits, which are set by the vibrational energy $\chi$. Measuring these energies in units of $\chi$, we characterize the three paths together by their net weight
	\begin{equation}
		\mathcal W=\frac{\chi^2}{d_{210}d_{120}}+\frac{\chi^2}{d_{210}d_{111}}+\frac{\chi^2}{d_{201}d_{111}},
		\label{eq:B_W}
	\end{equation}
	so that $V_3\simeq\sqrt{12}\,\mathcal W J_1^2J_2/\chi^2$. The three terms are the contributions of paths 1, 2, and 3, and $\mathcal W$ is the result of their interference. A path contributes positively when its two intermediate states lie on the same side of the resonant pair and negatively when they lie on opposite sides. To leading order, the coupling $V_3$ within $P$ is generated by the conversion Hamiltonian $H_{\rm conv}=\zeta_3(\hat c_1^{\dagger2}\hat c_2^\dagger\hat a^3+{\rm H.c.})$ of Eq.~\eqref{eq:Hconv}. The bosonic factors $\sqrt{3!}$ for removing the photons and $\sqrt{2!}$ for creating two subensemble-1 excitations give $\langle021|H_{\rm conv}|300\rangle=\sqrt{12}\,\zeta_3$. We use $\zeta_3$ for the leading-order conversion strength and $V_3$ for the full matrix element of Eq.~\eqref{eq:B_Heff}, which also contains higher-order virtual exchange and the losses of the eliminated states. The leading-order relation $V_3\simeq\sqrt{12}\,\zeta_3$ gives the conversion strength
	\begin{equation}
		\zeta_3\simeq\mathcal W\,\frac{J_1^2J_2}{\chi^2}.
		\label{eq:B_zeta3}
	\end{equation}
	This shows that $\mathcal W$ is the conversion strength in units of $J_1^2J_2/\chi^2$, the scale set by three exchanges and two intermediate states detuned by energies of order $\chi$.
	
	Without dissipation, Eq.~\eqref{eq:A_energy} gives the energies of the intermediate states and of the composite,
	\begin{equation}
		\begin{aligned}
			d_{210}&=\Delta_x-\Delta_a-\chi,\\
			d_{201}&=\Delta_x-\Delta_a-4\chi,\\
			d_{120}&=2(\Delta_x-\Delta_a)-2\chi+U_1,\\
			d_{111}&=2(\Delta_x-\Delta_a)-5\chi+U_{12},\\
			d_{021}&=3(\Delta_x-\Delta_a)-6\chi+U_1+2U_{12}.
		\end{aligned}
		\label{eq:B_dnu}
	\end{equation}
	Each conversion of a photon into an emitter excitation contributes $\Delta_x-\Delta_a$, each excitation of subensemble 1 or 2 contributes its polaron shift $-\chi$ or $-4\chi$, and the interaction energies $U_1$ and $U_{12}$ enter only when two excitations occupy subensemble 1 or both subensembles are excited.
	
	The three paths differ only in when the subensemble-2 excitation is created. Creating it costs $d_{201}$ from the three-photon state, but $d_{111}-d_{210}=d_{201}+U_{12}$ or $d_{021}-d_{120}=d_{201}+2U_{12}$ when one or two subensemble-1 excitations are already present. At the bare resonance $d_{021}=0$, the net weight~\eqref{eq:B_W} then takes the form
	\begin{equation}
		\mathcal W=\frac{\chi^2U_{12}\left(U_1+2U_{12}\right)}{d_{210}d_{120}d_{111}d_{201}},
		\label{eq:B_WU12}
	\end{equation}
	which yields Eq.~\eqref{eq:V3_shared}. This shows that the cross interaction $U_{12}$, which measures how the cost of the subensemble-2 excitation depends on the occupation of subensemble 1, controls whether the three paths add or cancel.
	
	\emph{Shared vibration, $s=-1$.} Here $U_{12}=4\chi$, and at the compensated detunings $\Delta_x-\Delta_a=O(J^2/\chi)$ is negligible on the scale of $\chi$. Equation~\eqref{eq:B_dnu} then gives $d_{210}=d_{111}=-\chi$, $d_{201}=d_{120}=-4\chi$, and $d_{021}=0$, so that all intermediate states lie below the resonant pair: $|210\rangle$ and $|111\rangle$, with $|Q|=1$, by $\chi$, and $|201\rangle$ and $|120\rangle$, with $|Q|=2$, by $4\chi$. Path~1 visits $|210\rangle$ and $|120\rangle$ and contributes $1/4$ to $\mathcal W$. Path~2 visits the two $|Q|=1$ states $|210\rangle$ and $|111\rangle$ and contributes $1$. Path~3 visits $|201\rangle$ and $|111\rangle$ and contributes $1/4$. Because all intermediate states lie on the same side of the resonant pair, the three contributions have the same sign and add to
	\begin{equation}
		\mathcal W=\frac14+1+\frac14=\frac32
		\label{eq:B_W_minus}
	\end{equation}
	[Fig.~\ref{fig:paths}(a)], with path 2, which avoids the $|Q|=2$ states, dominating. Equation~\eqref{eq:B_zeta3} then gives the conversion strength $\zeta_3\simeq3J_1^2J_2/(2\chi^2)$ of Eq.~\eqref{eq:Hconv}, or $V_3\simeq3\sqrt3\,J_1^2J_2/\chi^2$. Five-exchange paths and the compensated detunings correct this value at relative order $J^2/\chi^2$. For $s=-1$, the shared vibration thus provides both the resonance and a constructive conversion coupling.
	
	\emph{Shared vibration, $s=+1$.} Here $U_{12}=-4\chi$, and the same detunings give $d_{111}=-9\chi$, corresponding to $Q_{111}=3$, and $d_{021}=-16\chi$, while the other three energies are unchanged. Path~1 is therefore unaffected and contributes $1/4$ to $\mathcal W$, whereas paths 2 and 3 pass through $|111\rangle$, now $9\chi$ below $|300\rangle$, and contribute only $1/9$ and $1/36$. All intermediate states still lie below $|300\rangle$, so that the three contributions add to $\mathcal W=1/4+1/9+1/36=7/18$ [Fig.~\ref{fig:paths}(b)], $7/27$ of the value for $s=-1$. Since the composite lies $16\chi$ below $|300\rangle$, this coupling admixes it only with amplitude $V_3/(16\chi)=O(J^3/\chi^3)$ and shifts $|300\rangle$ by $V_3^2/(16\chi)=O(J^6/\chi^5)$, far below the residual two-photon shift of Eq.~\eqref{eq:B_Kerr}. The composite therefore remains an off-resonant state, as in Appendix~\ref{app:lamb}. At the detunings that bring the $s=-1$ composite into resonance, the $s=+1$ configuration thus has a finite coupling but no resonance.
	
	\emph{Independent vibrations.} Here $U_{12}=0$, and the bare resonance $d_{021}=0$ requires $\Delta_x-\Delta_a=8\chi/3$. Equation~\eqref{eq:B_dnu} then gives $d_{210}=5\chi/3$, $d_{120}=4\chi/3$, $d_{111}=\chi/3$, and $d_{201}=-4\chi/3$, so that the intermediate states now lie on both sides of the resonant pair. Path~1 visits $|210\rangle$ and $|120\rangle$, both above the resonant pair, and contributes $9/20$ to $\mathcal W$. Path~2 visits $|210\rangle$ and $|111\rangle$, again both above, and contributes $9/5$. Path~3 visits $|201\rangle$ below and $|111\rangle$ above the resonant pair and therefore contributes with the opposite sign, $-9/4$. This contribution cancels the other two,
	\begin{equation}
		\mathcal W=\frac{9}{20}+\frac95-\frac94=0
		\label{eq:B_W_ind}
	\end{equation}
	[Fig.~\ref{fig:paths}(c)]. The cancellation follows from $U_{12}=0$ alone: the subensemble-2 excitation then costs $d_{201}$ irrespective of when it is created, and the three paths combine into an amplitude proportional to $d_{021}$, which vanishes at resonance [Eq.~\eqref{eq:B_Wr_ind} with $r=2$]. With independent vibrations, the composite can thus be tuned into resonance but has no leading-order coupling to $|300\rangle$. At the compensated detunings of Eq.~\eqref{eq:B_detunings_ind}, the shifts of the energies $d_\nu$ by $O(J^2/\chi)$ and the finite linewidths from cavity and emitter losses leave a residual coupling smaller than the shared-vibration value by factors of order $(J/\chi)^2$ and $\kappa_a/\chi$, respectively, and the full weak-driving calculation of Fig.~\ref{fig:threephoton}(a) shows no resonant dip.
	
	\emph{Integer displacement ratios.} For $|g_2|=rg_1$ with integer $r\geq2$, in the dispersive regime $\sqrt{r}\,J\ll\chi$, where the bosonically enhanced exchange strengths remain small compared with the displacement energies, Eq.~\eqref{eq:A_energy} gives the polaron shift $-r^2\chi$ of a subensemble-2 excitation and the cross interaction $U_{12}=-2sr\chi$, and the shared vibration has $Q_\nu=n_1+srn_2$. For $s=-1$, the state $|0,r,1\rangle$ is undisplaced and, at the bare resonance $\Delta_x=\Delta_a$, where the Lamb shifts are not yet included, degenerate with the $(r+1)$-photon state $|r{+}1,0,0\rangle$; we measure the energies $d_\nu$ of the intermediate states from this pair. Converting the $r+1$ photons takes $r+1$ exchanges, and the paths differ only in the step $k=1,\ldots,r+1$ at which the subensemble-2 excitation is created. As in Eq.~\eqref{eq:B_threepaths}, the bosonic factors $\sqrt{(r+1)!}$ for removing the photons and $\sqrt{r!}$ for creating the subensemble-1 excitations are common to all paths, as is the matrix element $\langle0,r,1|H_{\rm conv}^{(r+1)}|r{+}1,0,0\rangle=\sqrt{(r+1)!\,r!}\,\zeta_{r+1}$ of the conversion Hamiltonian $H_{\rm conv}^{(r+1)}=\zeta_{r+1}(\hat c_1^{\dagger r}\hat c_2^\dagger\hat a^{r+1}+{\rm H.c.})$. Equation~\eqref{eq:B_pathsum} then gives the conversion strength in terms of the net weight $\mathcal W_{r+1}$ of the $r+1$ paths,
	\begin{equation}
		\zeta_{r+1}\simeq\mathcal W_{r+1}\frac{J_1^rJ_2}{\chi^r},\qquad
		\mathcal W_{r+1}=\sum_{k=1}^{r+1}\prod_{\nu\in k}\Big(-\frac{\chi}{d_\nu}\Big),
		\label{eq:B_Wr}
	\end{equation}
	where the product runs over the $r$ intermediate states of path $k$. For $r=2$, $\mathcal W_3$ is the net weight $\mathcal W$ of Eq.~\eqref{eq:B_W}, and Eq.~\eqref{eq:B_Wr} reduces to Eq.~\eqref{eq:B_zeta3}.
	
	For the shared vibration with $s=-1$, the detunings that compensate the Lamb shifts differ by $\Delta_x-\Delta_a=O(J^2/\chi)$, which is negligible on the scale of $\chi$, as for $r=2$. To leading order, the energy of each intermediate state is therefore its displacement energy, $d_\nu\simeq-\chi Q_\nu^2$, so that all intermediate states lie below the resonant pair and every path contributes positively. On path $k$, the first $k-1$ intermediate states contain $j=1,\ldots,k-1$ excitations of subensemble 1 and have $Q_\nu=j$, and the remaining $r-k+1$ contain the subensemble-2 excitation and $j=k-1,\ldots,r-1$ excitations of subensemble 1 and have $Q_\nu=j-r$. Path $k$ thus contributes $[(k-1)!\,(r-k+1)!]^{-2}$, and the paths sum to
	\begin{equation}
		\mathcal W_{r+1}=\sum_{k=1}^{r+1}\frac{1}{[(k-1)!\,(r-k+1)!]^2}=\frac{1}{(r!)^2}\sum_{j=0}^{r}\binom{r}{j}^2=\frac{1}{(r!)^2}\binom{2r}{r},
		\label{eq:B_Wr_shared}
	\end{equation}
	where the last step is Vandermonde's identity. With Eq.~\eqref{eq:B_Wr}, this gives Eq.~\eqref{eq:zeta_r}, and $r=2$ recovers $\mathcal W=3/2$ of Eq.~\eqref{eq:B_W_minus}; as for $r=2$, the compensated detunings and paths with further exchanges correct this value at relative order $J^2/\chi^2$. The paths that create the subensemble-2 excitation near the middle of the sequence dominate, because they avoid the most strongly displaced states, as path 2 does for $r=2$. For a ratio $|g_2|/g_1=r+\epsilon$ near the integer $r$, the intermediate states that contain the subensemble-2 excitation have $Q_\nu=j-r-\epsilon$, so that every path still contributes positively and $\mathcal W_{r+1}$ changes linearly with $\epsilon$; for $r=2$, Eq.~\eqref{eq:B_W} gives $\mathcal W\simeq3/2-11\epsilon/4$. For $s=+1$, the same state has $Q=2r$ and lies $4r^2\chi$ below the $(r+1)$-photon state, so that the conversion acts without a resonance. This shows that the shared vibration with $s=-1$ couples the $(r+1)$-photon state to an undisplaced composite for every integer $r$, with a strength that decreases with the photon number through the factor $J/\chi$ per exchange and through $\mathcal W_{r+1}$.
	
	For two independent vibrations, the displacement energies are $-\chi(n_1^2+r^2n_2^2)$, and the bare resonance of $|0,r,1\rangle$ with the $(r+1)$-photon state requires $(r+1)(\Delta_x-\Delta_a)=2r^2\chi$. Creating the subensemble-2 excitation then costs $d_{r,0,1}=\Delta_x-\Delta_a-r^2\chi$ at every step, irrespective of the occupation of subensemble 1, so that $d_{r-j,j,1}=d_{r+1-j,j,0}+d_{r,0,1}$ for $j=0,\ldots,r$, with $d_{r+1,0,0}=0$; for $r\geq2$, no intermediate state is resonant at these detunings. Consecutive paths differ in a single intermediate state, and adding the paths in the order of increasing $k$ gives, for $m=1,\ldots,r$,
	\begin{equation}
		\sum_{k=1}^{m}\prod_{\nu\in k}\Big(-\frac{\chi}{d_\nu}\Big)=-\frac{\chi}{d_{r,0,1}}\prod_{j=1}^{m-1}\Big(-\frac{\chi}{d_{r+1-j,j,0}}\Big)\prod_{j=m}^{r-1}\Big(-\frac{\chi}{d_{r-j,j,1}}\Big).
		\label{eq:B_Wr_partial}
	\end{equation}
	Adding the last path, which creates the subensemble-2 excitation at the final step, yields
	\begin{equation}
		\mathcal W_{r+1}=\frac{d_{0,r,1}}{d_{r,0,1}}\prod_{j=1}^{r}\Big(-\frac{\chi}{d_{r+1-j,j,0}}\Big).
		\label{eq:B_Wr_ind}
	\end{equation}
	The product describes the virtual conversion of $r$ photons into subensemble 1, and the prefactor is the energy of the composite in units of the cost of the subensemble-2 excitation. The sum therefore vanishes at the bare resonance $d_{0,r,1}=0$, for any polaron shifts and interaction energies within subensemble 1. This shows that, for every integer displacement ratio, independent vibrations can be tuned to the same resonance but give no leading-order conversion, and that the conversion of Eq.~\eqref{eq:zeta_r} is generated by the cross interaction of the shared vibration.
	
	\subsection{Residual two-photon shift}
	\label{app:kerr}
	
	The virtual exchanges that generate the three-photon conversion also make the dressed cavity ladder anharmonic. Here we show that this anharmonicity appears only at fourth order in the exchange couplings, one order higher than the conversion, so that the conversion suppresses $g^{(3)}(0)$ while leaving $g^{(2)}(0)$ close to unity.
	
	The same elimination in the two-excitation manifold, with $P=\{|200\rangle\}$, determines how far the dressed two-photon level deviates from twice the dressed one-photon energy. This residual two-photon shift,
	\begin{equation}
		\mathcal E_{200}^{(s)}\equiv\langle200|H_{\rm eff}^{P}|200\rangle-2\mathcal E_{100},
		\label{eq:B_E200def}
	\end{equation}
	is the anharmonicity of the dressed cavity ladder. At the compensated detunings, where $\mathcal E_{100}=0$, it equals the effective energy of $|200\rangle$. Each photon pair contributes it once, so that $|300\rangle$, which contains three pairs, has the effective energy $3\mathcal E_{100}+3\mathcal E_{200}^{(s)}$.
	
	Without the interaction energies $U_1$, $U_2$, and $U_{12}$, the cavity and the two emitter modes form a linear system, whose dressed ladder is harmonic at every order in the exchange. The residual shift is therefore set by these interactions, which act only when two virtual emitter excitations are present at the same time, that is, in the four-exchange processes of Eq.~\eqref{eq:B_pathsum}. At the compensated detunings, the interaction-dependent part of these processes gives
	\begin{equation}
		\begin{aligned}
			\mathcal E_{200}^{(-)}&=-\frac{1}{\chi^3}\left(J_1^4-\frac52J_1^2J_2^2+\frac{J_2^4}{64}\right),\\
			\mathcal E_{200}^{(+)}&=-\frac{1}{\chi^3}\left(J_1^4+\frac{5}{18}J_1^2J_2^2+\frac{J_2^4}{64}\right),
		\end{aligned}
		\label{eq:B_Kerr}
	\end{equation}
	with corrections of order $J^6/\chi^5$. The three terms correspond to the two virtual excitations residing both in subensemble 1, one in each subensemble, and both in subensemble 2. Only the mixed term depends on $s$, through the energy of the mixed state $|011\rangle$, which has $Q_{011}=-1$ for $s=-1$ and $Q_{011}=3$ for $s=+1$.
	
	Since the three-photon conversion [Eq.~\eqref{eq:B_paths}] requires one exchange fewer, $\mathcal E_{200}^{(s)}/\zeta_3=O(J/\chi)$ for both configurations: in the dispersive regime, the residual two-photon shift is parametrically weaker than the conversion.
	
\begin{figure}
		\centering
		\includegraphics[width=0.75\linewidth]{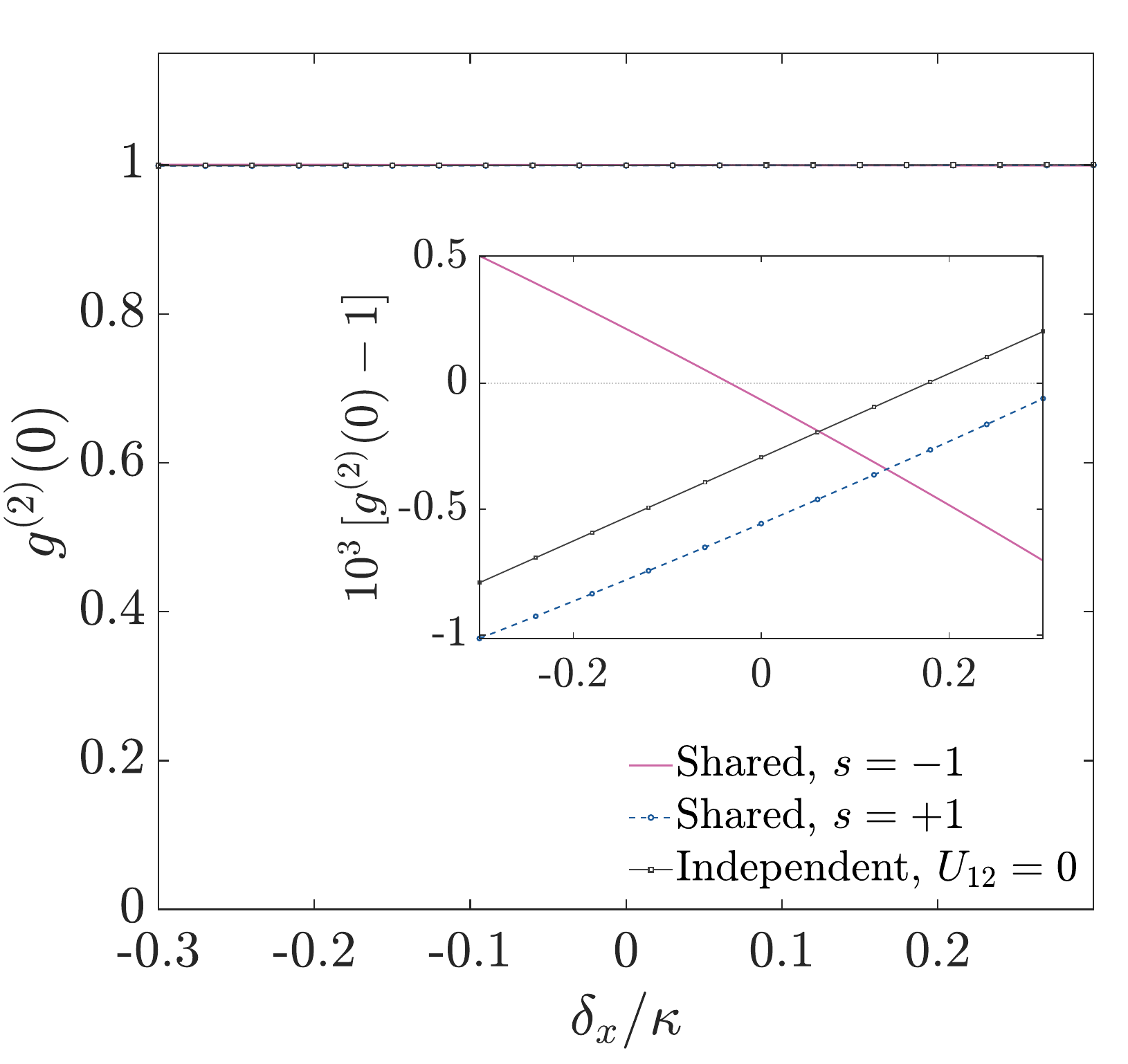}
		\caption{
			Two-photon correlation $g^{(2)}(0)$ versus the deviation $\delta_x$ of the emitter detuning from its compensated value, for the three cases and detunings of Fig.~\ref{fig:threephoton}(a), from the weak-driving amplitudes of Eq.~\eqref{eq:A_recursion}. Inset: deviation $g^{(2)}(0)-1$ on an enlarged scale. Parameters as in Fig.~\ref{fig:threephoton}(a).
		}
		\label{fig:g2}
	\end{figure}

	The same interactions determine how far $g^{(2)}(0)$ departs from unity. Without $U_1$, $U_2$, and $U_{12}$, the coherently driven cavity and emitter modes remain in a coherent state, for which $g^{(2)}(0)=1$ at every order in the exchange and for any losses. The interactions modify the two-photon amplitude only through the four-exchange processes above. Since the dressed photon is driven on resonance and broadened by $\kappa_a$, the real shift $\mathcal E_{200}^{(s)}$ changes $g^{(2)}(0)$ only at second order, whereas the losses of the doubly excited intermediate states and the drive of their emitter components change it at first order, with a relative weight of order $(J/\chi)^4$. Hence
	\begin{equation}
		|g^{(2)}(0)-1|\sim\Big(\frac{J}{\chi}\Big)^4+\Big(\frac{\mathcal E_{200}^{(s)}}{2\kappa_a}\Big)^2,
		\label{eq:B_g2dev}
	\end{equation}
	where $|\mathcal E_{200}^{(s)}|/\kappa_a$ is of order $(J/\chi)^2(\eta_3^{\rm cav})^{1/2}$, with the cavity-limited cooperativity of Eq.~\eqref{eq:B_eta3cav}. Independent vibrations carry a residual shift of the same order, set by $U_1$ and $U_2$. Figure~\ref{fig:g2} shows that, for the three cases and detunings of Fig.~\ref{fig:threephoton}(a), $g^{(2)}(0)$ stays at unity on the scale of the three-photon dip and deviates from it by about $10^{-3}$, in agreement with Eq.~\eqref{eq:B_g2dev}. The deviation varies linearly with the emitter detuning. For the shared vibration, a finite $\delta_x$ detunes the dressed photon from the drive by ${\rm Re}\,\mathcal E_{100}\simeq(J_1^2+J_2^2/16)\,\delta_x/\chi^2$ [Eq.~\eqref{eq:B_E100}], which combines with the residual shift into the correction $-{\rm Re}\,\mathcal E_{100}\,\mathcal E_{200}^{(s)}/\kappa_a^2$ to $g^{(2)}(0)$. The slope thus has the sign of $-\mathcal E_{200}^{(s)}$ in Eq.~\eqref{eq:B_Kerr}, negative for $s=-1$ and positive for $s=+1$; independent vibrations, whose residual shift is negative, show a positive slope as well. This shows that the resonant conversion suppresses $g^{(3)}(0)$ while the four-exchange processes leave $g^{(2)}(0)\simeq1$ for both configurations.
	
	\subsection{Photon correlations and the three-photon cooperativity}
	\label{app:corr}
	
	The resonant conversion of Appendix~\ref{app:zeta3} suppresses three-photon events only if the coupling $V_3$ competes with the losses of the three-photon state and of the composite. Here we restore cavity and emitter losses and show that, for $s=-1$ at resonance, $g^{(3)}(0)$ is set by a single three-photon cooperativity, which compares $V_3^2$ with the product of these two decay rates and, for narrow emitter lines, saturates at a value set by the cavity loss.
	
	Losses enter through the complex energies $\alpha=\Delta_a-i\kappa_a$ and $\beta_\mu$ of Eq.~\eqref{eq:A_energy}. The effective energies then acquire imaginary parts, both from the losses of the retained states themselves and from those of the off-resonant states, which the virtual processes of Eq.~\eqref{eq:B_pathsum} transfer to them: $-{\rm Im}\,\mathcal E_{100}$ is the decay rate of the dressed photon, and $\Gamma_{021}\equiv-{\rm Im}\,\mathcal E_{021}$ that of the composite. With losses, $\mathcal E_{100}$ coincides with the dressed complex cavity detuning $\Lambda_1$ of Eq.~\eqref{eq:A_C1}. Collecting Eqs.~\eqref{eq:B_E100}, \eqref{eq:B_E021}, \eqref{eq:B_zeta3}, and \eqref{eq:B_Kerr}, the effective Hamiltonian of Eq.~\eqref{eq:B_Heff} takes, to leading order in the dispersive expansion, the operator form
	\begin{equation}
		H_{\rm eff}\simeq\mathcal E_{100}\,\hat a^\dagger\hat a+\tfrac12\mathcal E_{200}^{(s)}\,\hat a^{\dagger2}\hat a^2+\mathcal E_{021}\,|021\rangle\langle021|+H_{\rm conv},
		\label{eq:B_Hdressed}
	\end{equation}
	which describes a dressed cavity with a weak anharmonicity, coupled to the composite by the three-photon conversion of Eq.~\eqref{eq:Hconv}. All off-resonant states lie at energies of order $\chi$ from the drive and are populated only virtually; their effect is contained in the dressed parameters $\mathcal E_{100}$, $\mathcal E_{200}^{(s)}$, $\mathcal E_{021}$, and $\zeta_3$. Adding the drive $\Omega(\hat a^\dagger+\hat a)$, the weak-driving equations~\eqref{eq:A_recursion} for the steady-state amplitudes $C_\nu$ become
	\begin{equation}
		\begin{aligned}
			\mathcal E_{100}C_{100}&=-\Omega,\\
			\big(2\mathcal E_{100}+\mathcal E_{200}^{(s)}\big)C_{200}&=-\sqrt2\,\Omega C_{100},\\
			\big(3\mathcal E_{100}+3\mathcal E_{200}^{(s)}\big)C_{300}+V_3C_{021}&=-\sqrt3\,\Omega C_{200},\\
			V_3C_{300}+\mathcal E_{021}C_{021}&=0,
		\end{aligned}
		\label{eq:B_ladder}
	\end{equation}
	where $\mathcal E_{100}$ and $\mathcal E_{021}$ include the cavity and emitter losses, $\mathcal E_{200}^{(s)}$ is the lossless residual shift of Eq.~\eqref{eq:B_Kerr}, and $V_3$ is the conversion coupling of Eq.~\eqref{eq:B_paths}. The full weak-driving equations~\eqref{eq:A_recursion}, with which Fig.~\ref{fig:threephoton}(b) compares this description, contain two further effects of relative order $(J/\chi)^2$. The drive also reaches the off-resonant states from the emitter components of the lower manifolds, for example $|010\rangle\to|110\rangle$, from which they couple back to the photonic states $|n00\rangle$, and, with losses, the residual shift acquires an imaginary part of order $(J/\chi)^2\kappa_a$, which describes a loss of photon pairs that differs from twice the single-photon loss. Both effects are present without the interactions, where the system remains in a coherent state, and therefore cancel in the photon correlations [Eq.~\eqref{eq:B_g2dev}]; Eqs.~\eqref{eq:B_ladder} and \eqref{eq:B_g23} accordingly contain the lossless, interaction-induced shift of Eq.~\eqref{eq:B_Kerr}.
	
	Apart from the residual shift of Eq.~\eqref{eq:B_Kerr}, the three cases differ only in the composite. For $s=-1$, it is resonant, $d_{021}=0$ in Eq.~\eqref{eq:B_dnu}, and coupled to $|300\rangle$ with the net weight $\mathcal W=3/2$ of Eq.~\eqref{eq:B_W_minus}. For $s=+1$, it lies at $d_{021}=-16\chi$ and only shifts $|300\rangle$ by $V_3^2/(16\chi)=O(J^6/\chi^5)$, so that the last equation and the term $V_3C_{021}$ drop out. For independent vibrations, it is resonant, but $\mathcal W=0$ [Eq.~\eqref{eq:B_W_ind}].
	
	The composite is fed only through $|300\rangle$. Eliminating $C_{021}$ adds the self-energy $-V_3^2/\mathcal E_{021}$ to the three-photon energy, so that the composite acts as a lossy reservoir for the three-photon amplitude. Equation~\eqref{eq:A_gn} then gives
	\begin{equation}
		g_s^{(2)}(0)\simeq\frac{4|\mathcal E_{100}|^2}{\big|2\mathcal E_{100}+\mathcal E_{200}^{(s)}\big|^2},\qquad
		g_s^{(3)}(0)\simeq\frac{36|\mathcal E_{100}|^4}{\big|2\mathcal E_{100}+\mathcal E_{200}^{(s)}\big|^2\big|3\mathcal E_{100}+3\mathcal E_{200}^{(s)}-V_3^2/\mathcal E_{021}\big|^2}.
		\label{eq:B_g23}
	\end{equation}
	For a harmonic ladder without conversion, $\mathcal E_{200}^{(s)}=V_3=0$, and both correlations equal unity. The residual shift enters both correlations, whereas the composite enters only $g^{(3)}(0)$. This shows that the two-photon statistics probe the anharmonicity of the dressed cavity, while the three-photon statistics additionally probe the conversion.
	
	For $s=-1$ at resonance, ${\rm Re}\,\mathcal E_{100}={\rm Re}\,\mathcal E_{021}=0$, which the detunings of Eq.~\eqref{eq:resonance_compensation} achieve up to the corrections of order $J^4/\chi^3$ discussed below Eq.~\eqref{eq:B_E021}. With losses, $V_3$ acquires an imaginary part smaller than its real part by a factor of order $\kappa_a/\chi$, which we neglect here and in Eq.~\eqref{eq:C_g3exact}. For $|\mathcal E_{200}^{(-)}|\ll\kappa_a$, Eq.~\eqref{eq:B_g23} then reduces to
	\begin{equation}
		g_-^{(3)}(0)\simeq\frac{1}{(1+\eta_3)^2},\qquad
		\eta_3=\frac{V_3^2}{3\,(-{\rm Im}\,\mathcal E_{100})\,\Gamma_{021}}.
		\label{eq:B_control}
	\end{equation}
	The three-photon cooperativity $\eta_3$ is the ratio of the squared conversion coupling to the product of the decay rate $-3\,{\rm Im}\,\mathcal E_{100}$ of the three-photon state and the decay rate $\Gamma_{021}$ of the composite. The depth of the dip is thus controlled by $\eta_3$ alone, and $g_-^{(3)}(0)$ decreases monotonically as $\eta_3$ grows. For $s=+1$ and for independent vibrations, the composite term $V_3^2/\mathcal E_{021}$ in Eq.~\eqref{eq:B_g23} is negligible, and $g^{(3)}(0)$ shows no resonant dip.
	
	For $s=-1$, to leading order, $-{\rm Im}\,\mathcal E_{100}\simeq\kappa_a$, and the composite decays both through its three emitter excitations, at the rate $3\kappa_x$, and through its virtual admixture of $|111\rangle$ and $|120\rangle$, each containing one cavity photon. With the exchange strengths and energy separations of Eq.~\eqref{eq:B_E021}, these states are admixed with probabilities $2J_1^2/\chi^2$ and $J_2^2/(16\chi^2)$, and both decay at the rate $\kappa_a+2\kappa_x$, so that
	\begin{equation}
		\Gamma_{021}\simeq3\kappa_x+(\kappa_a+2\kappa_x)\frac{2J_1^2+J_2^2/16}{\chi^2}.
		\label{eq:B_bT}
	\end{equation}
	With the leading-order relation $V_3\simeq\sqrt{12}\,\zeta_3$, the cooperativity becomes
	\begin{equation}
		\eta_3\simeq\frac{4\zeta_3^2}{\kappa_a\Gamma_{021}}=\frac{4\zeta_3^2}{\kappa_a}\left[3\kappa_x+(\kappa_a+2\kappa_x)\frac{2J_1^2+J_2^2/16}{\chi^2}\right]^{-1}.
		\label{eq:B_eta3}
	\end{equation}
	The composite decay rate in Eq.~\eqref{eq:B_eta3} falls as the emitter line narrows, so that the cooperativity grows and saturates for $\kappa_x\to0$ at the cavity-limited value
	\begin{equation}
		\eta_3^{\rm cav}\simeq\frac{4\zeta_3^2\chi^2}{\kappa_a^2\left(2J_1^2+J_2^2/16\right)}.
		\label{eq:B_eta3cav}
	\end{equation}
	A narrow emitter line therefore increases the cooperativity only up to $\eta_3^{\rm cav}$, so that $g_-^{(3)}(0)$ cannot fall below $(1+\eta_3^{\rm cav})^{-2}$. Since $J_1/\chi$ and $J_2/\chi$ are small in the dispersive regime, a cooperativity of order unity requires the polaron shift $\chi$ to exceed the cavity linewidth by a factor of order $(\chi/J)^2$ and the geometric mean $\sqrt{\kappa_a\kappa_x}$ of the cavity and emitter decay rates by a factor of order $(\chi/J)^3$.
	
	Reaching this floor requires the composite to be resonant on the scale of its decay rate, which for $\kappa_x\to0$ is of order $\kappa_aJ^2/\chi^2$. The detunings of Eq.~\eqref{eq:resonance_compensation} leave the composite off resonance by the corrections of order $J^4/\chi^3$ discussed below Eq.~\eqref{eq:B_E021}, and the ratio of this offset to the decay rate is of order $(\eta_3^{\rm cav})^{1/2}$. For a deep dip, $\eta_3^{\rm cav}\gg1$, the leading-order detunings therefore no longer keep the composite resonant, and in Fig.~\ref{fig:threephoton}(b) the emitter detuning is determined from the resonance condition $\mathcal E_{021}=0$ of Eq.~\eqref{eq:B_Heff} without losses, with the cavity detuning of Eq.~\eqref{eq:resonance_compensation}. The solid and dotted lines of Fig.~\ref{fig:threephoton}(b) evaluate Eq.~\eqref{eq:B_control} with $V_3$, $\mathcal E_{100}$, and $\mathcal E_{021}$ obtained from Eq.~\eqref{eq:B_Heff} including losses, the dotted line in the limit $\kappa_x\to0$.
	
	\section{Dephasing of the three-photon resonance}
	\label{app:dephasing}
	
	In this Appendix, we analyze how the dephasing terms of Eq.~\eqref{eq:ME} modify the three-photon resonance of Appendix~\ref{app:dispersive}. It answers three questions in turn: how vibrational and emitter dephasing act on the resonant pair, $|300\rangle$ and $|021\rangle$ (Appendix~\ref{app:deph_pair}), how they change $g^{(3)}(0)$ (Appendix~\ref{app:deph_corr}), and up to which dephasing rates the dip survives (Appendix~\ref{app:deph_tol}).
	
	We consider the shared vibration with $s=-1$ at resonance, in the weak-driving limit and in the dispersive regime $J_\mu\ll\chi$ of Appendix~\ref{app:dispersive}, with the cavity and emitter losses included as in Appendix~\ref{app:corr}. The two dephasing terms of Eq.~\eqref{eq:ME} randomize relative phases without changing the bare Fock-state populations. Together they damp the coherence between two Fock states $|\nu\rangle$ and $|\nu'\rangle$ at the rate
	\begin{equation}
		\Gamma_{\nu\nu'}=\frac{\Gamma_v}{2}(Q_\nu-Q_{\nu'})^2+\sum_\mu\phi_\mu(n_{\mu,\nu}-n_{\mu,\nu'})^2,
		\label{eq:C_damping}
	\end{equation}
	where $n_{\mu,\nu}$ is the number of excitations of subensemble $\mu$ in $|\nu\rangle$. The vibrational part vanishes when both states displace the vibration equally, and the emitter part when both contain the same excitations.
	
	\subsection{Dephasing of the resonant pair}
	\label{app:deph_pair}
	
	Without dephasing, the composite $|021\rangle$ loses its phase only through its decay, at the rate $\Gamma_{021}$ of Eq.~\eqref{eq:B_bT}. Here we show that dephasing separates the two: emitter dephasing adds only to the rate at which the composite loses its phase, whereas vibrational dephasing reaches the resonant pair only through the small components of the intermediate states that virtual exchange admixes into it, at second order in $J/\chi$, and adds to its decay rate.
	
	\emph{Emitter dephasing.} The composite $|021\rangle$ contains two excitations of subensemble 1 and one of subensemble 2, and the photonic states $|n00\rangle$ contain none. Its coherence with every photonic state is therefore damped at the rate $4\phi_1+\phi_2$, while the coherences among the photonic states remain undamped. Emitter dephasing leaves the bare populations unchanged, so that it destroys the phase of the composite and preserves its population.
	
	\emph{Vibrational dephasing.} For $s=-1$, the photonic states and the composite all have $Q=0$, and the vibrational term leaves the coherences among them undamped. It acts on the resonant pair only through the intermediate states of Fig.~\ref{fig:paths}(a). A single cavity--emitter exchange couples each resonant state to intermediate states that lie at energies of order $\chi$ away, so that the resonant state acquires small components of these states, with probabilities given by the squared ratio of the exchange strength to the energy separation. These admixed components displace the vibration. The composite contains $|111\rangle$ and $|120\rangle$, which have $Q_{111}=-1$ and $Q_{120}=2$, with the probabilities $2J_1^2/\chi^2$ and $J_2^2/(16\chi^2)$ of Eq.~\eqref{eq:B_bT}. The three-photon state contains $|210\rangle$ and $|201\rangle$, which have $Q_{210}=1$ and $Q_{201}=-2$, with the probabilities $3J_1^2/\chi^2$ and $3J_2^2/(16\chi^2)$; for example, $|210\rangle$ is reached from $|300\rangle$ with the exchange strength $\sqrt3J_1$ and lies $\chi$ away, which gives $(\sqrt3J_1/\chi)^2$. Vibrational dephasing causes leakage from the dressed composite into the off-resonant states through its $|111\rangle$ and $|120\rangle$ admixtures, at the rate $\Gamma_v(2J_1^2+J_2^2/4)/\chi^2$, in which each intermediate state is weighted by its probability and the square of its displacement. These states lie at energies of order $\chi$ from the resonant pair, so that the leakage damps the amplitude of the composite at half this rate. This leakage acts on the resonant pair in the same way as a decay, although no excitation is lost. Each photon similarly contains $|010\rangle$ and $|001\rangle$, which have $Q_{010}=1$ and $Q_{001}=-2$, with the probabilities $J_1^2/\chi^2$ and $J_2^2/(16\chi^2)$, which adds $\Gamma_v(J_1^2+J_2^2/4)/(2\chi^2)$ to the cavity decay rate $\kappa_a$; the three-photon state, which contains three times these admixtures, leaks at three times this rate. In what follows, $\kappa_a$ includes this contribution. The admixtures also give the composite and the three-photon state slightly different mean displacements, so that vibrational dephasing dephases them relative to each other at a rate of order $\Gamma_v(J/\chi)^4$; Eq.~\eqref{eq:C_fourth} gives this correction.
	
	Collecting these contributions, the composite decays at the rate $\Gamma_{021}$ and loses its phase relative to the photonic states at the rate $\tilde\Gamma_{021}$, with
	\begin{equation}
		\begin{aligned}
			\Gamma_{021}&\simeq3\kappa_x+(\kappa_a+2\kappa_x)\frac{2J_1^2+J_2^2/16}{\chi^2}+\frac{\Gamma_v}{2}\,\frac{2J_1^2+J_2^2/4}{\chi^2},\\
			\tilde\Gamma_{021}&\simeq\Gamma_{021}+4\phi_1+\phi_2.
		\end{aligned}
		\label{eq:C_rates}
	\end{equation}
	This shows that the decay of the composite and the loss of its phase become two distinct rates. Emitter dephasing enters only $\tilde\Gamma_{021}$, at full strength. Vibrational dephasing enters through the leakage at order $(J/\chi)^2$, which adds equally to both rates, because the displacement compensation that makes $|021\rangle$ resonant also leaves it undisplaced.
	
	\subsection{Three-photon correlations under dephasing}
	\label{app:deph_corr}
	
	Appendix~\ref{app:deph_pair} shows that, with dephasing, the composite decays at the rate $\Gamma_{021}$ but loses its phase at the larger rate $\tilde\Gamma_{021}$. Here we show how these two rates enter $g^{(3)}(0)$. Dephasing weakens the coherent conversion, which is governed by $\tilde\Gamma_{021}$, and turns part of the converted population into an incoherent population of the composite, which returns to $|300\rangle$ without interfering and thereby refills the dip. We describe the resonant pair and the photonic ladder with Eq.~\eqref{eq:B_ladder}, at resonance and for $|\mathcal E_{200}^{(-)}|\ll\kappa_a$, and obtain the steady state of this description in closed form.
	
	\emph{Coherent part.} The weak-driving amplitudes $C_\nu$ of Eq.~\eqref{eq:B_ladder} describe the coherences between the driven states and the vacuum, so that $|C_\nu|^2$ is the population of $|\nu\rangle$ that keeps a fixed phase relative to the drive. The coherence of the composite with the vacuum decays at the rate $\tilde\Gamma_{021}$, so that Eq.~\eqref{eq:B_ladder} holds with $\mathcal E_{100}=-i\kappa_a$ and $\mathcal E_{021}=-i\tilde\Gamma_{021}$. The composite amplitude follows the three-photon amplitude as $C_{021}=-iV_3C_{300}/\tilde\Gamma_{021}$, and the three-photon amplitude is reduced to
	\begin{equation}
		|C_{300}|^2=\frac{|C_{300}^{(0)}|^2}{(1+\tilde\eta_3)^2},\qquad
		\tilde\eta_3=\frac{4\zeta_3^2}{\kappa_a\tilde\Gamma_{021}},
		\label{eq:C_coherent}
	\end{equation}
	where $C_{300}^{(0)}$ is the three-photon amplitude without conversion. The coherent suppression is thus set by the cooperativity $\tilde\eta_3$, in which the composite enters through its phase-loss rate.
	
	\emph{Incoherent part.} With pure dephasing, the population of $|\nu\rangle$ in the density operator $\rho$ of Eq.~\eqref{eq:ME} exceeds $|C_\nu|^2$ by an incoherent part $\rho_\nu^{\rm inc}$, which has no fixed phase relative to the drive. Pure dephasing, at the rate $\tilde\Gamma_{021}-\Gamma_{021}=4\phi_1+\phi_2$, transfers the coherent composite population into this incoherent part at the rate $2(\tilde\Gamma_{021}-\Gamma_{021})|C_{021}|^2$. The incoherent composite population decays at the rate $2\Gamma_{021}$ and is converted back into $|300\rangle$ through the coherence between $|300\rangle$ and $|021\rangle$, which decays at the rate $3\kappa_a+\tilde\Gamma_{021}$. In the steady state, this coherence transfers population between the two states at the rate $\Gamma_{\rm conv}=2V_3^2/(3\kappa_a+\tilde\Gamma_{021})$, so that
	\begin{equation}
		\begin{aligned}
			0&=-6\kappa_a\rho_{300}^{\rm inc}+\Gamma_{\rm conv}\left(\rho_{021}^{\rm inc}-\rho_{300}^{\rm inc}\right),\\
			0&=2\left(\tilde\Gamma_{021}-\Gamma_{021}\right)|C_{021}|^2-2\Gamma_{021}\rho_{021}^{\rm inc}-\Gamma_{\rm conv}\left(\rho_{021}^{\rm inc}-\rho_{300}^{\rm inc}\right).
		\end{aligned}
		\label{eq:C_rateeq}
	\end{equation}
	The first line states that the returned photon triplets leave the cavity at the rate $6\kappa_a$ of a three-photon population. Since they carry no fixed phase, they add incoherently to $|C_{300}|^2$, and $g^{(3)}(0)=(|C_{300}|^2+\rho_{300}^{\rm inc})/|C_{300}^{(0)}|^2$.
	
	Solving Eqs.~\eqref{eq:C_coherent} and \eqref{eq:C_rateeq} gives
	\begin{equation}
		g_-^{(3)}(0)=\frac{(3\kappa_a)^2\left[\Gamma_{021}\tilde\Gamma_{021}(3\kappa_a+\tilde\Gamma_{021})+V_3^2(\tilde\Gamma_{021}-\Gamma_{021})\right]}{\left(3\kappa_a\tilde\Gamma_{021}+V_3^2\right)\left[3\kappa_a\Gamma_{021}(3\kappa_a+\tilde\Gamma_{021})+V_3^2\left(3\kappa_a+\Gamma_{021}\right)\right]}.
		\label{eq:C_g3exact}
	\end{equation}
	In the regime of interest, the composite is much narrower than the three-photon state, $\tilde\Gamma_{021}\ll3\kappa_a$, and Eq.~\eqref{eq:C_g3exact} reduces to
	\begin{equation}
		g_-^{(3)}(0)\simeq\frac{1+\eta_3-\tilde\eta_3}{(1+\eta_3)(1+\tilde\eta_3)},
		\label{eq:C_g3}
	\end{equation}
	where $\eta_3$ is the cooperativity of Eq.~\eqref{eq:B_eta3}, now evaluated with the decay rate $\Gamma_{021}$ of Eq.~\eqref{eq:C_rates}. Without dephasing, $\tilde\eta_3=\eta_3$ and Eq.~\eqref{eq:C_g3} reduces to Eq.~\eqref{eq:B_control}. For strong dephasing, Eq.~\eqref{eq:C_g3exact} approaches unity. This shows that dephasing acts on the three-photon statistics in two ways. It weakens the coherent conversion, which reduces the cooperativity from $\eta_3$ to $\tilde\eta_3$, so that less of the three-photon amplitude is cancelled. It also turns a fraction $(\tilde\Gamma_{021}-\Gamma_{021})/\tilde\Gamma_{021}$ of the converted population, the share of pure dephasing in the phase loss of the composite, into the incoherent population of Eq.~\eqref{eq:C_rateeq}, which returns to $|300\rangle$ and refills the dip.
	
	\subsection{Dephasing scales of the dip and higher-order corrections}
	\label{app:deph_tol}
	
	We now determine the dephasing rates at which the dip of Eq.~\eqref{eq:C_g3} is appreciably filled, and estimate the corrections beyond the rates of Eq.~\eqref{eq:C_rates}. Emitter dephasing fills the dip once $4\phi_1+\phi_2$ reaches $\Gamma_{021}/\eta_3$, whereas vibrational dephasing must exceed this scale by a factor of order $\eta_3(\chi/J)^2$.
	
	\emph{Emitter dephasing.} For a deep dip, $\eta_3\gg1$, and for $4\phi_1+\phi_2\ll\Gamma_{021}$, Eq.~\eqref{eq:C_g3} becomes
	\begin{equation}
		g_-^{(3)}(0)\simeq\frac{1}{(1+\eta_3)^2}\left[1+\eta_3\,\frac{4\phi_1+\phi_2}{\Gamma_{021}}\right].
		\label{eq:C_g3deep}
	\end{equation}
	The prefactor is the coherent residual $(1+\eta_3)^{-2}$ of Eq.~\eqref{eq:B_control}, and the second term in the bracket is the incoherent return of photon triplets relative to this residual. This ratio grows with the cooperativity, because the population converted into the composite exceeds the residual three-photon population by a factor proportional to $\eta_3$. The three-photon correlation therefore doubles when $4\phi_1+\phi_2$ reaches $\Gamma_{021}/\eta_3=\kappa_a\Gamma_{021}^2/(4\zeta_3^2)$, which lies well below $\Gamma_{021}$ for $\eta_3\gg1$. A deeper dip is thus more sensitive to emitter dephasing.
	
	\emph{Vibrational dephasing.} To the order of Eq.~\eqref{eq:C_rates}, vibrational dephasing changes only $\Gamma_{021}$ and $\kappa_a$, so that the dip keeps the form $(1+\eta_3)^{-2}$ of Eq.~\eqref{eq:B_control} with these rates. It becomes appreciably shallower when the leakage $\Gamma_v(2J_1^2+J_2^2/4)/(2\chi^2)$ becomes comparable to the decay rate of the composite without vibrational dephasing, Eq.~\eqref{eq:B_bT}. In the cavity limit, this decay rate is the cavity loss $\kappa_a(2J_1^2+J_2^2/16)/\chi^2$ inherited through the same intermediate states $|111\rangle$ and $|120\rangle$, and the condition becomes $\Gamma_v\sim\kappa_a$, independent of $J/\chi$ to leading order. Vibrational dephasing and cavity loss then compete on equal footing, because both reach the composite only through its admixed components.
	
	The dephasing rates at which the two mechanisms appreciably fill the dip thus differ by a factor of order $\eta_3(\chi/J)^2$, given to leading order in $J/\chi$ by $2\eta_3\chi^2/(2J_1^2+J_2^2/4)$, with $\eta_3$ evaluated without vibrational dephasing. This shows that the displacement compensation protects the dip against the dephasing of the shared vibration: vibrational dephasing must exceed emitter dephasing by this factor to have the same effect.
	
	\emph{Fourth-order vibrational dephasing.} The admixtures of Appendix~\ref{app:deph_pair} also give the two resonant states small mean displacements. Weighting the displacements of the intermediate states by their probabilities, the composite $|021\rangle$ acquires the mean displacement $-(2J_1^2-J_2^2/8)/\chi^2$, and the three-photon state $|300\rangle$ acquires $3(J_1^2-J_2^2/8)/\chi^2$. The two differ by $(5J_1^2-J_2^2/2)/\chi^2$, so that vibrational dephasing damps the coherence between the two resonant states at the additional rate
	\begin{equation}
		\Gamma_v^{(4)}=\frac{\Gamma_v}{2\chi^4}\Big(5J_1^2-\frac{J_2^2}{2}\Big)^2,
		\label{eq:C_fourth}
	\end{equation}
	where the superscript indicates the fourth order in $J/\chi$. We estimate its effect by treating this relative dephasing of the resonant pair in the same way as emitter dephasing and adding it to $4\phi_1+\phi_2$ in Eq.~\eqref{eq:C_g3deep}, so that vibrational dephasing also produces a small incoherent return; for the couplings of Fig.~\ref{fig:dephasing}, this estimate reproduces the master-equation results. Relative to the leakage, it changes the vibrational scale by a fraction of order $\eta_3(J/\chi)^2$. This fraction is small for moderate dips and reaches order unity only for the deepest dips. There, the vibrational scale is set by this fourth-order dephasing, and it still exceeds the emitter scale by a factor of order $(\chi/J)^4$, because this dephasing is suppressed by $(J/\chi)^4$. The analytic lines of Fig.~\ref{fig:dephasing} use Eq.~\eqref{eq:C_g3exact} with the rates of Eq.~\eqref{eq:C_rates}, Eq.~\eqref{eq:C_fourth} added to $\tilde\Gamma_{021}$, and the dressed parameters of Fig.~\ref{fig:threephoton}(b); the master-equation results are obtained from the excitation-resolved steady state of Eq.~\eqref{eq:ME} in the weak-driving limit.

	\begin{center}
	{\bf{ACKNOWLEDGMENTS}}
\end{center}
This work was supported by the Natural Science Foundation of Jilin Province (Grant No.~20240101013JC), the National Natural Science Foundation of China (Grant No.~12375020), the Talents Introduction Project of Anhui Science and Technology University (Grant Nos.~DQYJ202301 and DQYJ202006), the University-level Research Development Fund Project (Grant No.~FZ20250115), and the Scientific Research Foundation for Higher Education Institutions of Anhui Province (Grant No.~2024AH040057).

\end{document}